\documentclass[12pt, a4paper]{article}
    \usepackage{graphicx}
    \begin{document}

    \title{Cosmological Phase Transitions\footnote{Invited lecture at the $3^{rd}$ Summer School on
    Condensed Matter Research, 7-14 August 2004, Zuoz, Switzerland.}}

    \author{Norbert Straumann\\
        Institute for Theoretical Physics University of Zurich,\\
        CH--8057 Zurich, Switzerland}

    \maketitle

    \begin{abstract}
    In this lecture at a school for condensed matter physicists, I begin with basic concepts and tools
    for investigating phase transitions in quantum field theory. The very different roles
    of global and gauge symmetries in phase transitions will be elucidated. Among the
    important applications of the basic theory the thermodynamics of the electroweak
    transition is  treated in detail, and the implications for baryogenesis will be
    discussed. The status of lattice simulations of the transition from a high temperature
    quark-gluon plasma to confined hadronic matter with spontaneously broken chiral
    symmetry is also briefly reviewed. The various changes of the free energy density in
    cosmological phase transitions reinforce the cosmological constant problem. In the
    last part of the lecture I shall address this profound mystery of present day physics.
    \end{abstract}

    \section{Introduction}

During the evolution of the very early Universe there have been at least two phase
transitions. The electroweak theory predicts that at about 100 GeV there was a transition
from a `symmetric' high temperature phase with `massless' gauge bosons to the Higgs phase,
in which the $SU(2)\times U(1)$ gauge symmetry is `spontaneously broken' and all the masses
in the model are generated. A detailed understanding of the nature and dynamics of this
transition is a very difficult task, and a lot of quantitative analytical and numerical
studies have been performed over the years. One of the motivations of these studies has
been that non-perturbative effects might have changed the baryon asymmetry in the phase
transition.

Quantum chromodynamics (QCD) predicts that at about 200 MeV there was a transition from a
quark-gluon plasma to a confinement phase with no free quarks and gluons. At about the same
energy we expect that the global chiral symmetry of QCD with massless fermions is
spontaneously broken by the formation of a quark pair condensate. These transitions have
been studied extensively in the framework of lattice gauge theory. It is unclear whether
they have interesting cosmological implications.

In the framework of unified extensions of the standard model other potentially very
interesting phase transitions are expected. In some cases these are accompanied by the
formation of topological defects (domain walls, strings, monopoles and textures). This can
be cosmologically interesting or dangerous. For example, in a second order phase transition
of a grand unified theory (GUT) one expects generically that a completely unacceptable
abundance of magnetic monopoles would be formed. This so-called monopole problem was one of
the early motivations of inflationary cosmology. The first inflationary model by A.Guth
(`old inflation') was based on a first order phase transition, and made use of the enormous
energy density of the `false vacuum' for the unbroken phase.

These introductory remarks indicate the potential importance of phase
transitions\footnote{I shall sometimes use the term phase transition in a loose sense, even
if a detailed analysis may later show that the transition is a crossover, rather than a
\textit{bona fide} phase transition.} in cosmology. In this evening lecture I can, of
course, treat only some selected aspects of the subject. Since this is a school for
condensed matter physicists, I shall begin with foundational material for investigating
phase transitions in quantum field theory. A central tool of most studies is the path
integral representation of the partition function. In this connection I want to make some
critical remarks about the widespread use of non-convex effective potentials, especially in
the older literature, that disturbs me very much. I also want to explain why the roles of
global and gauge symmetries in phase transitions are very different. In particular, a
\textit{gauge symmetry can not be spontaneously broken} (Elitzur theorem). As an important
application of the basic theory, I shall review existing work on the thermodynamics of the
electroweak phase transition, and its role in baryogenesis. We shall continue with a few
remarks about the current status of lattice simulations of the QCD phase transitions. The
subject of cosmic topological defects that might have been formed in phase transitions will
only be touched.

The various changes of the free energy density in cosmological phase transitions reinforces
the \textit{cosmological constant problem}, a deep mystery of present day physics. In the
last part of my talk I shall address this topical issue, and then summarize the current
evidence for a dominant component of what people (unfortunately) call \textit{Dark Energy}
in our Universe. Whether this is a tiny remainder of an originally huge vacuum energy in
the very early Universe remains open, but observations may eventually settle this issue.

\section{Phase Transitions in Quantum\\ Field Theory}

Let us first consider a quantum field theory without local symmetries. The partition
function can (formally) be represented as an Euclidean path (functional) integral. If
$S[\phi]$ denotes the Euclidean action functional of a (multi-component) scalar field
$\phi$, say, and if we include the coupling to an external source $J(x)$, this
reads\footnote{We consider here only the canonical ensemble. For grand canonical ensembles
one has to replace in the action $\partial_4$ by $\partial_4-\mu$, where $\mu$ is the
chemical potential of the field.}
\begin{equation}
 Z[J] = \int_{\beta} \exp\{-S[\phi]+(\phi,J)\}~[d\phi],
\end{equation}
where the integration extends over fields which are periodic in Euclidean time $\tau$, with
periodicity $\beta=1/kT$. For textbook treatments see \cite{Be} and \cite{Kap}.

Several remarks on this formula are in order. In one dimension, i.e. for quantum mechanics,
it has -- suitably interpreted -- a precise meaning and can be rigorously derived for a
large class of interactions. In four dimensions this is rarely the case without an
ultraviolet cutoff. An exception are free fields (Gaussian probability measures on spaces
of distributions). In perturbation theory, renormalization theory leads -- for
renormalizable field theories -- order by order to meaningful expressions. In some cases,
for example for QCD, we expect that the continuum limit of lattice regularizations exists
and is physically sound. This is, however, not the case even for simple scalar models
(triviality theorem \cite{AF}). It is remarkable that this Euclidean formulation in terms
of path integrals establishes such a close connection with classical (!) statistical
mechanics (models of magnetism). For the study of the thermodynamics of quantum field
theories and other issues one can work entirely in the Euclidean framework. An explicit
continuation back to Minkowski space is not needed.

An important quantity is $(-\beta)$ times the free Helmholz energy (Schwinger functional):
\begin{equation}
 W[J]=\ln~Z[J],
\end{equation}
that generates the truncated correlation functions. The \textit{effective action} (Gibbs
potential) $\Gamma$ is the \textit{generalized} Legendre (\textit{Legendre-Fenchel})
transform of $W[J]$:
\begin{equation}
\Gamma[\Phi]=\sup_{J} \{(\Phi,J)-W[J]\}.
\end{equation}
Expectation values of observables $A[\phi]$ are given by
\begin{equation}
 \langle A\rangle_{J}=Z[J]^{-1}\int A[\phi]\exp\{-S[\phi]+(\phi,J)\}~[d\phi].
\end{equation}
$W[J]$ is convex in $J$. For a lattice regularization this follows simply from H\"{o}lder's
inequality (see footnote below). Formally, this is implied by
\[ \int f(x)\frac{\delta^2W[J]}{\delta J(x)\delta J(y)}f(y)d^4xd^4y=
\Bigl\langle\Bigl(\phi(f)-\langle\phi(f)\rangle\Bigr)^2\Bigr\rangle\geq0,\] for every real
test function. As a Legendre transform, $\Gamma[\Phi]$ is also convex. For a homogeneous
source $J$ the \textit{effective potential} $V_{eff}(\Phi)$ is, by definition, the free
Gibbs energy density.

Consider, for illustration, the simple example of an `action' without a kinetic energy term
\[S[\phi]=\mathcal{U}[\phi]=\int U(\phi)~d^4x.\]
For $T=0$ and a constant source $J$, the free energy density $w_0=W_0/\Omega_4$ in a finite
region of volume $\Omega_4$ of a lattice regularization is given by a one-dimensional
integral:
\begin{equation}
w_0(J)=\frac{1}{\Omega_4}\ln\int e^{\Omega_4(J\varphi-U(\varphi))} d\varphi.
\end{equation}
In the limit $\Omega_4\rightarrow \infty$ the Laplace approximation becomes exact, hence
\begin{equation}
w_0\rightarrow L(U),
\end{equation}
where $L$ denotes the Legendre-Fenchel transformation. This agrees (see Appendix A) with
$L$ of the convex hull of $U$, and since $L(L(U))$= convex hull of $U$, we obtain the
result that the Gibbs free energy density is for a constant $\Phi$ equal to the
\textit{convex hull of $U$}. In Fig. 1 we show for a non-convex $U$ the transforms $L(U)$
and $L(L(U))$.
\begin{figure}
\begin{center}
\includegraphics[height=0.25\textheight]{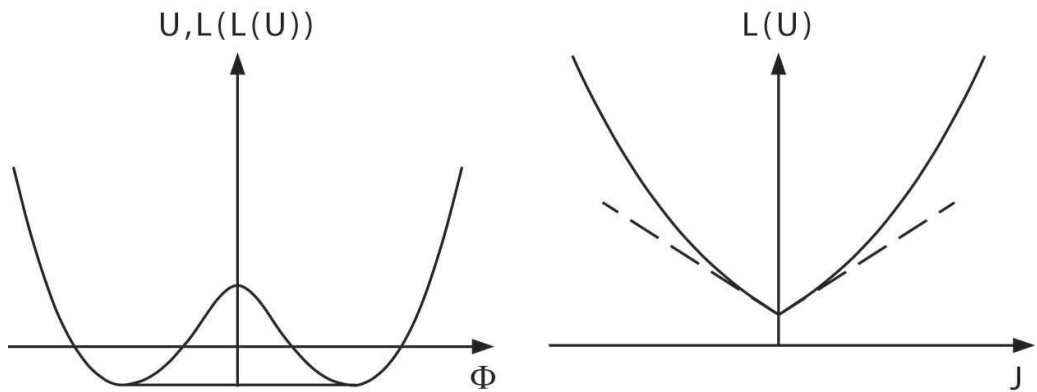}
\caption{Graphs of a non-convex potential $U$, together with its first and second
Legendre-Fenchel transforms.}
\label{Fig-1}
\end{center}
\end{figure}
$L(U)$ is still the free energy density for $T>0$. (Note that for $T>0$ the volume
$\Omega_4$ is $\beta$ times the three-volume $\Omega_3$.)

\subsubsection*{Phase transitions}

In Figs. 2 and 3 we show the typical shapes of effective potentials for first and second
order phase transitions. Their Legendre-Fenchel transformations are the Helmholtz free
energy densities. The reader is invited to sketch these. For instance, the slope of $w(J)$
corresponding to Fig. 2 develops for $T\downarrow T_c$ and $J=0$ a jump (discontinuity in
the `magnetization'), while for Fig. 3 the `magnetization' vanishes at $T\geq T_c$ and
starts to grow below $T_c$.

\begin{figure}
\begin{center}
\includegraphics[height=0.4\textheight]{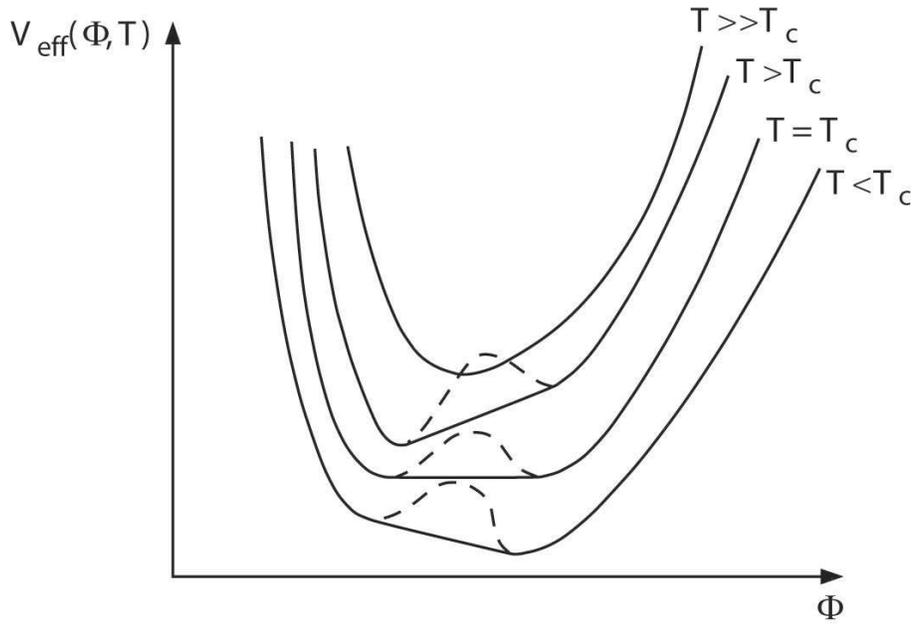}
\caption{Effective potential for a \textit{first order} phase transition. At the critical
temperature the minimum is degenerate (horizontal line corresponds to a phase mixture). The
order parameter makes a jump between the pure phases.} \label{Fig-2}
\end{center}
\end{figure}

\begin{figure}
\begin{center}
\includegraphics[height=0.25\textheight]{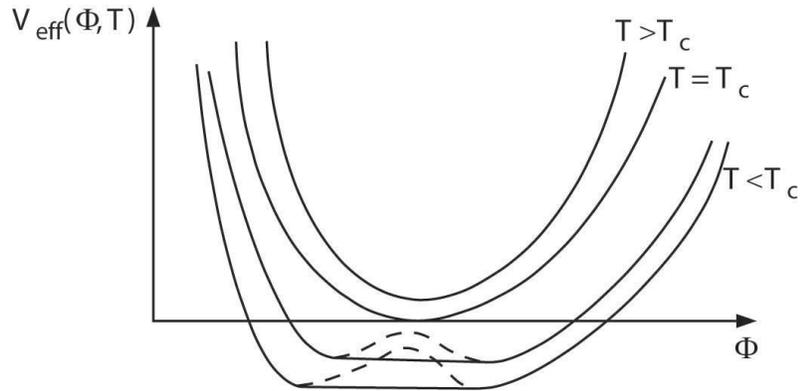}
\caption{Effective potential for a \textit{second order} phase transition. For the critical
temperature the second derivative vanishes at the minimum. The horizontal straight line
below $T_c$ shrinks to a point at $T_c$.}
\label{Fig-3}
\end{center}
\end{figure}

\subsubsection*{Footnote: Convexity of $W[J]$}

The convexity of $W[J]$ holds even in a finite volume, and not only in the thermodynamic
limit. (Actually, $W$ is then even \textit{strictly} convex). The lattice regularized
partition sum is then an integral of the form
\[ Z(\mbox{\boldmath$\alpha$})=\int \exp\Bigl(\mbox{\boldmath$\alpha\cdot f$}(x)\Bigr)d\mu(x), \]
where \mbox{\boldmath$\alpha$}=$(\alpha_1,...,\alpha_k)$ and \mbox{\boldmath$\alpha\cdot
f$}$(x)=\sum\alpha_i f_i(x)$. We claim that for any $\lambda$ with $0<\lambda<1$
\[ \ln Z\Bigl(\lambda\mbox{\boldmath$\alpha$}+(1-\lambda)\mbox{\boldmath$\beta$}\Bigr)\leq \lambda\ln
Z(\mbox{\boldmath$\alpha$})+(1-\lambda)\ln Z(\mbox{\boldmath$\beta$}). \]

This follows from the following application of H\"{o}lder's inequality
\[ \|FG\|_1\leq\|F\|_p\|G\|_q,~~~\frac{1}{p}+\frac{1}{q}=1~~(p>1),\]
for $F=e^{\lambda f},~G=e^{(1-\lambda)g},~p=1/\lambda,~q=1/(1-\lambda),~e^f\in L^1(\mu)$.

Usually, the convexity of $\ln Z $ is established by considering the quadratic form
\[ \frac{\partial^2\ln Z}{\partial\alpha_i\partial\alpha_j}\xi_i\xi_j= \langle (f_{\xi}-\langle
f_{\xi}\rangle)^2\rangle \geq 0, \] where $ f_{\xi}:=\sum_{i} f_{i}\xi_{i}$, and the
angular bracket denotes the expectation value with respect to the probability measure
$Z(\mbox{\boldmath$\alpha$})^{-1}\exp\Bigl(\mbox{\boldmath$\alpha\cdot f$}\Bigr) d\mu$.

This argument shows that $\ln Z$ is convex as a function of any parameter that appears
\textit{linearly } in the exponential under the integral sign in $Z$.

\subsection{Mean field approximation}

Countless expressions and plots in books and articles violate the basic convexity property,
usually as a result of some approximate calculations that are analytically extended to
regions where they are no more valid. Even in some really good books that happens already
for the mean field approximation. For illustration let me explain what is not done
properly.

Consider a one component neutral scalar field with Euclidean action in $D$ dimensions
\begin{equation}
S[\phi]=\int \Bigl[\frac{1}{2}(\partial \phi)^2 + U(\phi)\Bigr]~d^Dx \equiv
T[\phi]+\mathcal{U}[\phi].
\end{equation}
To have well-defined expressions interpret all the formulae in the lattice regularization
for finite (hypercube) subsets $\Lambda$ of the lattice. (Then $T[\phi]$ contains nearest
neighbor interactions.) Beside the action $S$ we consider an `approximate' action $S_0$ and
the corresponding probability measures
\begin{equation}
d\mu_0(\phi)=Z_o^{-1}e^{-S_0}[d\phi],~~~ Z_0=\int e^{-S_0}[d\phi]
\end{equation}
for all finite subsets $\Lambda$. The corresponding expectation values are denoted by
$\langle A \rangle_0$. The starting point is Jensen's inequality
\begin{equation}
\langle \exp A\rangle_0 \geq \exp (\langle A\rangle_0)
\end{equation}
for $A=-S+(\phi,J)+S_0$, and for $S_0$ we choose
\begin{equation}
S_0=-(\phi,K)+\mathcal{U}[\phi],
\end{equation}
where $K(x)$ is another external source. This gives, if $W_0=\ln~Z_0$,
\begin{equation}
W[J]\geq W_0 + \langle S_0 - S + (\phi,J)\rangle_0 =W_0[K] - \langle T[\phi]\rangle_0 +
((J-K),\langle\phi\rangle_0).
\end{equation}
For $\overline{\phi}:= \langle\phi\rangle_0$ we have
\begin{equation}
\overline{\phi}(x) = \frac{\partial W_0[K]}{\partial K(x)}.
\end{equation}
Since $\langle \phi(x)\phi(y)\rangle_0 = \langle\phi\rangle_0 \langle \phi(y)\rangle_0$, we
find
\begin{equation}
W[J]\geq \mathcal{W}[J,K]:= W_0[K]- T[\overline{\phi})] +((J-K),\overline{\phi}).
\end{equation}
The mean \textit{field free energy} $W_{mf}[J]$ is the supremum in $K$ of the
\textit{Landau functional } $\mathcal{W}[J,K]$, often also called the \textit{Landau free
energy}. Since $W_0[K]$ is convex, $K$ and $\overline{\phi}$ are according to (12) in one
to one correspondence. Hence we can write
\begin{equation}
W_{mf}[J]=\sup_{\overline{\phi}}\{(\overline{\phi},J)-(\Gamma_0[\overline{\phi}]+ T[
\overline{\phi}])\},
\end{equation}
where $\Gamma_0$ is the Legendre transform of $W_0$. This equation shows that $W_{mf}[J]$
is the Legendre-Fenchel transform of $\Gamma_0[\overline{\phi}]+T[\overline{\phi}]$, and is
therefore convex. Furthermore (see Appendix A), the mean free Gibbs potential is given by
\begin{equation}
\Gamma_{mf}[\Phi]= convex~ hull~\{\Gamma_0[\Phi]+T[\Phi]\}.
\end{equation}
It is  crucial that this is convex. Most authors work with the standard Legendre transform
and end up for $\Gamma_{mf}$ with the functional in the curly bracket, which is in general
not convex. (For a typical example, see Figs. 24.1, 2 in \cite{Z-J}.) Note also that the
Landau functional, expressed as a function of $J$ and $\overline{\phi}$, is given by the
curly bracket in (14)
\begin{equation}
\mathcal{W}[J,\overline{\phi}]=(J,\overline{\phi})-( \Gamma_0[\overline{\phi}]+T[
\overline{\phi}]).
\end{equation}
For $J=0$ its convex hull is according to (15) equal to $-\Gamma_{mf}$. We emphasize that
the Landau functional is in general \textit{not convex}.

As an exercise, one may work out everything for the Ising model.

\subsection{One-loop approximation}

The loop expansion of the Gibbs free energy $\Gamma[\Phi]$ is a much used systematic
approximation scheme. We recall here only the zero- and one-loop terms. These are obtained
with the steepest decent method.

The saddle points $\phi_0$ of the exponent in (1) satisfy the classical equation of motion
\[\left.\frac{\delta S}{\delta \phi}\right|_{\phi_0}=J.\]
Let us first assume that there is only one such saddle point for a given $J$. This is, for
instance, the case for an action of the form (7) for a convex potential $U(\phi)$, because
the previous equation then becomes
\[-\Delta \phi_0 + U'(\phi_0)= J, \]
which has at most one solution if $J(x)$ vanishes at infinity and $\phi_0$ is assumed to go
to zero at infinity. In the expansion
\[S[\phi] -(\phi,J)= S(\phi_0)-(\phi_0,J)+ \frac{1}{2}(\psi,A\psi)+...,\]
with $\psi:=\phi-\phi_0$, and
\begin{equation}
A(x,y)=\left.\frac{\delta^2 S}{\delta \phi(x)\delta\phi(y)}\right|_{\phi_0},
\end{equation}
we leave out the higher orders and arrive at
\begin{equation}
Z[J]\simeq \exp\{(\phi,J)-S[\phi_0]\}\int e^{-\frac{1}{2}(\psi,A\psi)} [d\psi].
\end{equation}
We interpret all the formulas in the lattice regularization (for a finite sublattice). The
leading order of $W[J]$ is given by the Laplace approximation
\begin{equation}
W_0[J]=(\phi_0,J)-S[\phi_0]=(LS)[J].
\end{equation}
Therefore, the leading order $\Gamma_0$ of $\Gamma$ is given by
\begin{equation}
\Gamma_0[\Phi]=S[\Phi]
\end{equation}
(`Landau approximation'). The one-loop correction is determined by the Gaussian integral
\[\int e^{-\frac{1}{2}(\psi,A\psi)}[d\psi]=\Bigl(\det \frac{A}{2\pi}\Bigr)^{-\frac{1}{2}}.\]
For $W[J]$ we arrive at
\begin{equation}
W[J]=W_0[J]+W_1[J]+...~,~~~W_1[J]=-\frac{1}{2}\ln\det\Bigl(\frac{A}{2\pi}\Bigr).
\end{equation}
For a given $J$ the conjugate variable $\Phi$ differs from the classical solution $\phi_0$
by a first order quantity. Up to second order we can thus replace $\phi_0$ by $\Phi$ in the
expression (21) for $W_1$. Therefore, the Legendre transformation (3)
\[\Gamma[\Phi]=\sup_{J} \{(\Phi,J)-LS[J]-W_1[\phi_0(J)]+ ...\}\]
is given by
\begin{equation}
\Gamma[\Phi]=\Gamma_0[\Phi]+\Gamma_1[\Phi]+...~,~~~\Gamma_1[\Phi]=\frac{1}{2}\ln\det\Bigl(\frac{A}{2\pi}\Bigr),
\end{equation}
where $A$ is now the expression (17) at $\phi=\Phi$.

If we set in (7) $U(\phi)=\frac{1}{2}m^2\phi^2+P(\phi)$, where $P(\phi)$ is a polynomial
(of degree $\leq 4$ for renormalizable interactions), then the 1-loop contribution of
$\Gamma(\Phi)$ is given by the following expression (up to $\Phi$ and $T$ independent
terms)
\begin{equation}
\Gamma_1(\beta,\Phi)=\frac{1}{2}\ln\det\left\{\frac{1}{2\pi}\left[-\Delta+m^2 +
P''(\Phi)\right]\right\}.
\end{equation}

The temperature dependence is contained in the Laplacian (periodicity in the Euclidean time
direction). The functional determinant can be worked out formally in the continuum theory
or from the lattice regularization, for which it is well-defined and can easily be computed
using Fourier transformation on a finite lattice. The details are given in Appendix B,
where we also show that the continuum limit of the temperature dependent part
$V_1(\beta,\Phi)$ of the effective potential (free energy density) exists, and is given by
\begin{equation}
V_1(\beta,\Phi)=\frac{1}{\beta}\int
\frac{d^3k}{(2\pi)^3}\ln\Bigl(1-e^{-\beta\Omega(k)}\Bigr),
\end{equation}
where
\begin{equation}
\Omega(k)=\sqrt{k^2+M^2},~~~M^2:=m^2+P''(\Phi).
\end{equation}
The reader is invited to generalize this to the grand canonical ensemble of a complex
scalar field.

The $T=0$ contribution requires renormalizations. As a starting point we use the lattice
regularization, derived in Appendix B. According to (63) we obtain in the thermodynamic
limit for an x-independent $\Phi$
\begin{equation}
\lim\frac{1}{\Omega_4}\ln\det\left[-\Delta+M^2\right]=\int_{BZ}\frac{d^4k}{(2\pi)^4}
\ln\Bigl[ M^2+\frac{2}{a^2}\sum_{\alpha} (1-\cos k_{\alpha} a)\Bigr],
\end{equation}
where the integral now extends over the four-dimensional Brillouin zone
$BZ=[-\frac{\pi}{a},\frac{\pi}{a}]^4$. In what follows it is simpler to use a spherical
cut-off $k^2\leq\Lambda^2$. In the continuum limit $a\rightarrow0$ we find for the
regularized one-loop effective potential $V_1(\Phi)$ at $T=0$, up to $\Phi$-independent
contributions,
\begin{equation}
V_1^\Lambda(\Phi)=\frac{1}{2}\int_{k^2\leq\Lambda^2}\ln\left[\frac{k^2+V_0''(\Phi)}{\mu^2}\right]
\frac{d^4k}{(2\pi)^4}
\end{equation}
$(V_0=U)$, where we have introduced an arbitrary mass scale $\mu$ (leading to a change of
$V_1$ which is independent of $\Phi$). This can be written as
\[ V_1(\Phi)=\frac{\mu^4}{32\pi^2}f(V_0''/\mu^2),\]
with
\[f(y)=\int_{0}^{\Lambda/\mu} \ln(x+y)x~dx.\]
We note that the third derivative of $f$ is given by
\[ f'''(y)=\int_{0}^{\Lambda/\mu}\frac{2x}{(x+y)^3}~dx\]
and converges for $\Lambda\rightarrow\infty$ to $1/y$. Hence, $f(y)=\frac{1}{2}y^2\ln(y)$ +
quadratic polynomial in $y$. So,
\[V_1(\Phi)=\frac{1}{64\pi^2}(V_0'')^2\ln\left(\frac{V_0''}{\mu^2}\right) + \mbox{quadratic polynomial
in} ~V_0''/\mu^2.\] The second order polynomial can be compensated by a renormalization of
the coupling constants in $V_0=U$. Note that we also have to renormalize a
$\Phi$-independent vacuum energy. This will become a crucial issue when gravity is included
(see Sect. 4).

After renormalization the effective potential for $T=0$ is, up to one loop, given by
\begin{equation}
V_{eff}(\Phi)=V_0(\Phi)+\frac{1}{64\pi^2}\Bigl[V_0''(\Phi)\Bigr]^2~\ln\Bigl[V_0''(\Phi)/\mu^2\Bigr].
\end{equation}

\textbf{Remark}. A change of the scale $\mu^2,~ \mu^2\rightarrow e^{t}\mu^2$, can be
absorbed by a change of the coupling constants in $V_0(\Phi;p(t))$, where the parameters
$p(t)$ satisfy, up to higher orders, the differential equation
\[ \frac{d}{dt}V_0(\Phi;p(t))=\frac{1}{64\pi^2}\Bigl[V_0''(\Phi:p(t))\Bigr]^2.\]
If
\[V_0(\varphi)=h+k\varphi+\frac{m^2}{2!}\varphi^2
+\frac{f}{3!}\varphi^3+\frac{g}{4!}\varphi^4,\] we obtain for $t=0$ the renormalization
group equation
\begin{equation}
\frac{dg}{dt}=\frac{3}{32\pi^2}g^2,~~~\mbox{etc.}
\end{equation}

\subsubsection*{Convex potentials}
For a convex potential like $U(\phi)=\frac{1}{2}m^2\phi^2+\frac{\lambda}{4}\phi^4$,
$M^2(\Phi)= m^2+3\lambda\Phi^2$ is positive, and the one-loop effective potential is
convex. The minimum occurs at $\Phi=0$, and in this symmetric phase the T-dependent
effective mass is
\begin{equation}
m^2(T)=m^2+\left.\frac{\partial^2V_1(\beta,\Phi)}{\partial\Phi^2}\right|_{\Phi=0}.
\end{equation}
The T-dependent term can be read off from (24)
\begin{equation}
\left.\frac{\partial^2V_1(\beta,\Phi)}{\partial\Phi^2}\right|_{\Phi=0}=3\lambda\int\frac{d^3k}{(2\pi)^3}
\frac{1}{\omega}\frac{1}{e^{\beta\omega}-1},~~~\omega=\sqrt{k^2+m^2}.
\end{equation}
For high temperatures we have
\begin{equation}
m^2(T)=m^2+\frac{1}{4}\lambda\Bigl[T^2-\frac{3}{4\pi}mT+\mathcal{O}(m^2\ln(m^2/T^2))\Bigr].
\end{equation}
For $m=0$ the scalar field acquires a mass $\sqrt{\lambda}T/2$ which plays the role of an
infrared cutoff.

\subsubsection*{Problems with non-convex potentials}

For a non-convex potential $U$ (with the `wrong' sign of the mass term) we can not simply
use the previous formulae, since $M^2(\Phi)$ becomes negative for small $\Phi$, whence
$V_1(\beta,\Phi)$ in (24) becomes \textit{complex}. This means that the one-loop potential
is not valid for small arguments, that are of particular interest in studies of the early
universe. What should one do in this situation? We shall address this question below.

For a rough estimate of the critical temperature for the transition to the symmetric phase
we may naively ignore this and use the leading corrections in (32):
\begin{equation}
m^2(T_c)=0\simeq m^2+\frac{1}{4}\lambda T_c^2.
\end{equation}
This gives (Kizhnit, Linde, Weinberg)
\begin{equation}
T_c^2\simeq\frac{(-4m^2)}{\lambda}~~~(m^2<0).
\end{equation}
For weak coupling this implies that $T_c^2\gg-m^2$, and we can therefore expect that (34)
is a good estimate.

The unrenormalized parameter $m^2$ in this equation can be replaced by the mass of the
physical particle (the ``$\sigma$ meson''). In lowest order, $m_\sigma^2=-2m^2$, hence
\begin{equation}
T_c^2=\frac{2m_\sigma^2}{\lambda}.
\end{equation}

We can, however, not ignore the encountered problems with the one-loop effective potential.
It is quite clear that the saddle point approximation can only hold near the minima. If
there are, for example, two similarly deep minima, as is expected close to the critical
temperature, one has to take both into account. As a recipe one might take the convex hull
of the sum of the saddle point approximations for the two minima. But this is just a
reasonable recipe. Another approach was studied in \cite{Cah}, where the standard
definition of the effective potential is modified by coupling the source to a
\textit{quadratic} polynomial of the scalar field. The new potential is not necessarily
convex as a function of the average value of $\phi$, but is convex as a function of the
composite field since this is the conjugate variable. The modified potential closely tracks
the usual one where the latter is real. Moreover, at finite temperatures it displays a
local minimum at $\Phi=0$. It is also worthwhile to note that the resulting critical
temperature in the one-loop approximation confirms the rough estimate (34).

\subsection{The constraint potential}

We consider now the free energy for a given spatial average of the order parameter. For
simplicity, we do this for a scalar field theory, since generalizations are obvious. As
before, the formulae below are well-defined in the lattice regularization. For a finite
region $\Lambda$ we sample the field configurations according to their mean field
\[\bar{\phi}=\frac{1}{\left|\Lambda\right|}\int \phi~d^Dx.\]
The probability distribution of $\bar{\phi}$ for a homogeneous source $J$ is given by
\begin{eqnarray*}
 dP(\bar{\phi}) & = & \left\langle\delta\Bigl(\bar{\phi}-\frac{1}{\left|\Lambda\right|}\int
\phi~d^Dx\Bigr)\right\rangle d\bar{\phi}\\
& = & e^{J\left|\Lambda\right|\bar{\phi}}d\bar{\phi}~Z_\Lambda(J)^{-1}\int
e^{-S_\Lambda[\phi]}
\delta\Bigl(\bar{\phi}-\frac{1}{\left|\Lambda\right|}\int \phi~d^Dx\Bigr)[d\phi]_\Lambda\\
& \equiv & \frac{1}{Z_{\Lambda}(J)} \exp\Bigl(\left|\Lambda\right|[J\bar{\phi}
-V_{c}(\bar{\phi})]\Bigr) d\bar{\phi}.
\end{eqnarray*}

The \textit{constraint potential } $V_c$ (which depends on $\Lambda$) is given in terms of
the integral in the last equation as
\begin{equation}
e^{-\left|\Lambda\right|V_{c}(\bar{\phi})}=\int e^{-S_\Lambda[\phi]}
\delta\Bigl(\bar{\phi}-\frac{1}{\left|\Lambda\right|}\int \phi~d^Dx\Bigr)[d\phi]_\Lambda.
\end{equation}
Since $dP$ is a probability measure, we have
\begin{equation}
Z_{\Lambda}(J)=\int e^{\left|\Lambda\right|[J\bar{\phi} -V_{c}(\bar{\phi})]} d\bar{\phi}.
\end{equation}
In other words, $W_\Lambda(j)$ is obtained from $V_c$ by a Laplace transformation. For
finite regions $V_c$ is usually not convex.

In the infinite volume limit the Laplace approximation again becomes exact if the saddle
point of $V_c-J\bar{\phi}$ is unique in this limit. Hence
\begin{equation}
\frac{1}{\left|\Lambda\right|}W_{\Lambda}(J)\rightarrow
w(J)=\sup_{\bar{\phi}}[J\bar{\phi}-V_{c}(\bar{\phi})]=(LV_c)(J),
\end{equation}
which implies that the effective potential is equal to the convex hull of the thermodynamic
limit of the constraint potential. Since it can be shown \cite{AW} that the latter is
convex (certainly if the saddle point is unique, but presumably in general), the two
potentials must be the same in this case.

Note that for $J=0$ the probability density $p(\bar{\phi})$ to find the system in a state
of `magnetization' $\bar{\phi}$ is given by
\[ p(\bar{\phi})=e^{-\left|\Lambda\right|V_{c}(\bar{\phi})}
\left[\int e^{-\left|\Lambda\right|V_{c}(\bar{\phi})}d\bar{\phi}\right]^{-1}.\]

The constraint effective potential has thus a very intuitive meaning. It has often been
studied via lattice simulations. For an example of an interesting recent application
(vacuum stability), see \cite{Ho}.

\subsection{Effective potential for gauge theories}

For gauge theories it is advisable to construct a gauge invariant effective potential, that
has a physical interpretation in terms of a gauge invariant order parameter.  In this
context it has to be stressed that for lattice regularizations of locally symmetric field
theories any \textit{local} quantity which is not gauge invariant, and has no gauge
invariant component (for precisions, see Appendix C), such as a  Higgs field, has a
vanishing expectation value at \textit{any temperature}. In this sense \textit{gauge
invariance can not be spontaneously broken} (Elitzur theorem \cite{Eli}). In perturbation
theory, where one expands about a Gaussian measure, this is not so because gauge invariance
is \textit{explicitly} broken by gauge fixing. In the lattice formulation one does,
however, not have to fix a gauge. (In practice, it may be useful to fix the gauge if the
ensemble of configurations is not very large.)

\subsubsection*{Lattice formulation of gauge theories}

The lattice regularization of gauge theories (Wilson) is of crucial importance in the
strong (confined) field regime, because there is no other way to perform numerical
calculations. But also from the conceptual point of view, lattice gauge theories are very
interesting. For systematic presentations, see \cite{MM} and \cite{Rot}.

In this discrete formulation of a gauge theory, belonging to a compact group $G$, the gauge
potential is replaced by a map of \textit{bonds} (ordered pairs $b$ of nearest neighbor
lattice points) into $G$: $b\mapsto g_b\in G$, with the property that $g_{x,y}=
g_{y,x}^{-1}$ for any bond $b=\langle x,y\rangle$. In order to formulate the analog of the
Yang-Mills action we need the notion of a \textit{plaquette}. This is a closed curve of
four bonds $P={\langle x,y\rangle, \langle y,z\rangle,\langle z,u\rangle\langle
u,x\rangle}$. To each of these we assign the group element $g_{\partial
P}=g_{x,y}g_{y,z}g_{z,u}g_{u,x}$, and for a given unitary character $\chi$ of the group $G$
the real number
\begin{equation}
S_P=\frac{1}{2}\Bigl(\chi(g_{\partial P)}+\chi^{*}(g_{\partial P})\Bigr).
\end{equation}
Note that $S_P$ is independent of the orientation of $P$. More generally, we can assign to
each closed path $\mathcal{C}$ on the lattice the product $g_{\mathcal{C}}$ of its $g_b$'s,
along one of the orientations, and then take its character $\chi(\mathcal{C})$. This number
is also independent of which site we begin the loop with. The action of a configuration
$\{g\}$ on a finite region $\Lambda$ of the lattice is taken to be
\begin{equation}
S_\Lambda({g})=-\sum_{P\subset\Lambda}S_P.
\end{equation}

The expectation value of an `observable' $A(\{g\})$ in $\Lambda$ is defined by
\begin{equation}
\langle A\rangle_\Lambda=\frac{1}{Z_\Lambda}\int
\exp(-S_\Lambda)A~d\mu_\Lambda,~~~Z_\Lambda=\int e^{-S_\Lambda}~d\mu_\Lambda,
\end{equation}
where $d\mu_\Lambda$ is the product of the normalized Haar measures for all bonds in
$\Lambda$.

A local gauge transformation is a map $x\mapsto \gamma_x\in G$. Under such a transformation
a gauge field configuration transforms as
\begin{equation}
g_{x,y}\mapsto \gamma_xg_{x,y}\gamma_y^{-1}~~~(b=\langle x,y\rangle).
\end{equation}
Clearly, $S_P$, the action and the measures $d\mu_\Lambda$ are invariant under local gauge
transformations.

Using the Baker-Hausdorff formula, it is easy to see that the naive formal continuum limit
leads to the Yang-Mills theory. There are good reasons to believe that the continuum limit
of a pure lattice gauge theory provides a consistent interacting Euclidean field theory.

Next, we add in a gauge invariant manner Higgs fields to the lattice model. To do this for
such a field $\phi(x)$, we have first to introduce a discrete form of the covariant
derivative. If this field transforms with respect to $G$ according to the unitary
representation $U(g)$, then the following definition has the right formal continuum limit
(we use the notation introduced in Appendix B)
\begin{equation}
D_\mu\phi(x)=\frac{1}{a}\Bigl[ U(g_{x,x+ae_\mu})\phi(x+ae_\mu)-\phi(x)\Bigr].
\end{equation}
The adjoint operator is
\begin{equation}
D_\mu^\dag \phi(x)=\frac{1}{a}\Bigl[ U(g_{x,x-ae_\mu})\phi(x-ae_\mu)-\phi(x)\Bigr].
\end{equation}
The corresponding gauge invariant Laplace operator (generalizing (57) in Appendix B) is
\begin{equation}
\Delta_{\{g\}}=-D_\mu^\dag D_\mu.
\end{equation}
The discrete Higgs part of the action is then in $D$ dimensions (dropping the
$\Lambda$-dependence)
\begin{equation}
S^H[\{g\},\phi]=\frac{1}{2}a^{D-2}\sum_{\langle x,y\rangle}\mid
U_{x,y}\phi(y)-\phi(x)\mid^2 +a^D\sum_{x}V(\phi(x)),
\end{equation}
where $U_{x,y}:=U(g_{x,y})$.

When fermionic fields are discretized in the most obvious way, the problem of `fermion
doubling' shows up. Early remedies to cure this involved an explicit breaking of chiral
symmetry of QCD (see, e.g., Chap. 4 of \cite{Rot}). Fortunately, there has been recently
important progress in solving this long-standing problem. Lattice formulations that
preserve chiral symmetry have been found, but in practice numerical simulations become much
harder. For a very readable recent review on these aspects, see \cite{CW}.

\subsubsection*{Continuum limit}

What we are really interested in is what the theory describes in the continuum limit. The
lattice is only used as an ultraviolet gauge invariant regularization, valid outside the
domain of perturbation theory.

Consider, for concreteness, a pure Yang-Mills lattice theory with a single
\textit{dimensionless} (bare) coupling constant $g$. The only dimensionful parameter in the
theory is the lattice spacing $a$. Therefore, a physical quantity with the dimension of a
mass, say, such as a particle mass $m_{phys}$, must be of the form
\[ m_{phys}=\frac{1}{a}f(g).\]
The theory allows us to compute $f(g)$, and thus ratios of masses. Clearly, in taking the
continuum limit $a\rightarrow 0$ we have to tune $g$ as a function of $a$ such that
$am_{phys}\rightarrow 0$. Since the inverse, $\xi=(am_{phys})^{-1}$, is a correlation
length, we have to approach -- in the language of statistical mechanics -- a
\textit{critical point} of the theory. The tuning of $g(a)$ is controlled by a
renormalization group equation of the form
\[a\frac{dg}{da}=-\beta_{LAT} (g).  \]
In lattice perturbation theory $\beta_{LAT}(g)$ can be computed as a power series in $g$
for small $g$:
\[ \beta(g)=-\beta_0 g^3 - \beta_1 g^5 + ... .\]
For $SU(N)$ one finds
\[ \beta_0 =\frac{N}{16\pi^2}\frac{11}{3},~etc.  \]
From this we find by integration
\[ g^2=-\frac{1}{\beta_0 \ln (a^2\Lambda_{LAT}^2)}+ ...~,\]
where $\Lambda_{LAT}$ is an integration constant. Inverting this one sees that
\[ a m_{phys}= const~\exp\Bigl(-\frac{1}{2\beta_0 g^2}\Bigr)\cdot (...),\]
showing the non-perturbative character of masses in the theory. Note that the bare coupling
constant $g(a)$ vanishes in the continuum limit.

The dimensionful \textit{lattice $\Lambda$-parameter}, $\Lambda_{LAT}$, provides a scale
which survives the continuum limit. This is remarkable, because classically the theory has
no scale. Through the process of renormalization we have introduced a mass scale into the
quantized theory, a mechanism that is often called \textit{dimensional transmutation}. This
renormalization procedure is in the lattice formulation no more mysterious than adjusting
the temperature toward the Curie point of a ferromagnet. At the same time these
considerations shows that the continuum limit has to be taken with care. They also show an
interesting point of contact with critical behavior, discussed in other lectures at this
school.

\subsubsection*{Elitzur's theorem for gauge theories}

For systems with global symmetries the phenomenon of spontaneous symmetry breaking (SSB) is
well-known. This is accompanied by a non-vanishing `magnetization'. At first sight one
expects something similar for gauge theories. However, Elitzur \cite{Eli} has shown that
\textit{local} quantities, like the bond variables $g_b$ or a Higgs field, which are not
gauge invariant, have always vanishing mean values. This is quite easy to prove, and is
also physically understandable. It is instructive to show this first for a lattice gauge
theory with gauge group $Z_2$, because this shows the contrast to the Ising model. In
Appendix C we give a general proof of Elitzur's theorem.

The bond variables will be denoted by $\sigma_b(=\pm 1$. The action for a finite region
$\Lambda$ of the lattice $Z^D$ is taken to be
\begin{equation}
S_\Lambda(\{\sigma\})=-\sum_{P\subset\Lambda}\sigma_{\partial
P}-h\sum_{b\subset\Lambda}\sigma_b,
\end{equation}
where $h$ is an external `field'. We could also include a Higgs field $\phi_i$,
transforming according to $\phi_i\mapsto \varepsilon_i\phi_i,~\varepsilon_i=\pm 1$.

Before continuing with this model, let us recall the SSB for the Ising model with its
global symmetry $Z_2$, consisting of the identity and the reflection
$\sigma_i\mapsto-\sigma_i$ for the Ising spins $\sigma_i$ of all lattice sites $i$. Above a
critical temperature $T_c$ there is only \textit{one} infinite-volume equilibrium (Gibbs)
state. However, for $T<T_c$ each translation invariant equilibrium state (probability
measure) $\mu^\beta$ is a convex linear combination of \textit{two different} extremal
states $\mu_{\pm}^\beta$. This means that $\mu^\beta$ describes a \textit{mixture} of two
\textit{pure phases}. The probability measures $\mu_{\pm}^\beta$ are weak limits of Gibbs
states on finite regions $\Lambda\subset Z^D$ with $\pm$ boundary conditions outside
$\Lambda$. Since they are different, they are not invariant under the symmetry group $Z_2$
of the interaction; the symmetry is \textit{spontaneously broken for these pure phases}.
Correspondingly, the spontaneous magnetizations
\begin{equation}
m_{\pm}(\beta)=\langle\sigma_ i\rangle_{\mu_{\pm}^\beta}
\end{equation}
do not vanish for $\beta>\beta_c$.

In sharp contrast to this situation, the mean value of $\langle\sigma_b\rangle$ does not
signal a symmetry breaking for the $Z_2$ lattice gauge model:

\vspace{1cm}

\textbf{Theorem (Elitzur)}. \textit{For the expectation value
$\langle\sigma_b\rangle_\Lambda(h)$ we have}
\begin{equation}
\lim_{h\downarrow 0}\langle\sigma_b\rangle=0~~~\mbox{\textit{uniformly in} $\Lambda$
\textit{and} $\beta$}.
\end{equation}
\textit{In particular, the thermodynamic limit of $\langle\sigma_b\rangle_\Lambda(h)$
vanishes for $h\downarrow 0$ (no spontaneous `magnetization')}. \vspace{1cm}

Before giving the simple proof, we remark that when a Higgs field $\phi_x$ is added, a
similar proof shows that
\begin{equation}
\langle\phi_x\rangle=0
\end{equation}
(Exercise). This does, however, not exclude a Higgs phase with an exponential fall off of
the truncated correlation function $\langle\sigma_{\partial P_1}\sigma_{\partial
P_2}\rangle_c$ for the gauge fields (mass generation), but (49) and (50) show that this is
not signaled by local gauge variant observables. This fact is well-known to people working
in lattice gauge theory, but is largely ignored outside this community. As order parameters
one may try to choose local gauge \textit{invariant} quantities, such as the norm of a
Higgs field.

\vspace{0.5cm}

\textit{Proof of Elitzur's theorem}.~ We can choose in (46) the bond variable $\sigma_{01}$
for the bond $b=\langle0,1\rangle$, and estimate in
\begin{equation}
\langle\sigma_{01}\rangle_\Lambda=\frac{\sum_{\{\sigma\}}\sigma_{01}\exp(\beta\sum_P
\sigma_{\partial P}+\beta h\sum_b\sigma_b)}{\sum_{\{\sigma\}}\exp(\beta\sum_P
\sigma_{\partial P}+\beta h\sum_b\sigma_b)}
\end{equation}
the numerator $N$ and the denominator $D$ separately.

Consider a gauge transformation $\sigma_{ij}'=\varepsilon_i\sigma_{ij}\varepsilon_j$, with
$\varepsilon_i=1$ for $i\neq 0$ and replace $\sigma_{ij}$ in $N$ and $D$ by
$\varepsilon_i\sigma_{ij}'\varepsilon_j$, dropping afterwards the prime of
$\sigma_{ij}~(\sigma_{ij}\mapsto\varepsilon_i\sigma_{ij}\varepsilon_j)$. This can be done
for $\varepsilon_0=\pm 1$. $D$ is equal to half the sum:
\[ D=\frac{1}{2}\sum{\varepsilon_0}\sum_{\{\sigma\}}\exp\Bigl(\beta\sum_P
\sigma_{\partial P}+\beta h\sum_b{}^\prime \sigma_b\Bigr)\exp\Bigl(\varepsilon_0\beta
h\sum_{j=1}^{2D} \sigma_{0j}\Bigr),\] where the prime of the sum means, that the bonds
$\langle 0,j\rangle$ must be excluded. Clearly, for $h>0$
\[D\geq \sum_{\{\sigma\}}\exp\Bigl(\beta\sum_P
\sigma_{\partial P}+\beta h\sum_b{}^\prime \sigma_b\Bigr)\frac{1}{2}e^{-2D\beta h}.\]
Similarly, we have for the numerator
\[|N|\leq \sum_{\{\sigma\}}\exp\Bigl(\beta\sum_P
\sigma_{\partial P}+\beta h\sum_b{}^\prime
\sigma_b\Bigr)\frac{1}{2}\Bigl|\sum{\varepsilon_0}\varepsilon_0\sigma_{01}\exp\Bigl(\varepsilon_0\beta
h\sum_{j=1}^{2D} \sigma_{0j}\Bigr)\Bigr|,\]
and thus
\[|N|\leq \sum_{\{\sigma\}}\exp\Bigl(\beta\sum_P
\sigma_{\partial P}+\beta h\sum_b{}^\prime \sigma_b\Bigr)\sinh(2D\beta h).\]

This gives the estimate
\begin{equation}
|\langle\sigma_{01}\rangle_\Lambda(h)|\leq 2e^{2D\beta h}\sinh(2D\beta h)\rightarrow 0
~~~\mbox{for $h\rightarrow 0$, uniformly in $\Lambda$ and $\beta$ }.
\end{equation}

What is the physical reason for this different behavior of models with local and global
symmetries? Consider once more the Ising model in the absence of an external field. At low
temperatures, the two regions in configuration space with opposite magnetizations
$\sigma_x$ and $-\sigma_x$ can only be connected by a path involving the creation of an
infinite interface which costs an infinite amount of energy. Therefore, the process cannot
occur spontaneously. Alternatively, one can make use of a small external field which is
switched off \textit{after} the thermodynamic limit is taken. In this procedure, one of the
two fundamental configurations $\{\sigma_x=+1\}$ and $\{\sigma_x=-1\}$ is energetically
eliminated and a non-vanishing magnetization is left. On the other hand, in a local gauge
theory one can perform symmetry transformations which act only non-trivially on a finite
set of basic variables on which a local `observable' depends. The system behaves therefore
similarly as quantum mechanical systems with  finite numbers of degrees of freedom. This
aspect is manifest in the proof given above, and becomes perhaps even more transparent in
the general proof of Elitzur's theorem given in Appendix C. Sometimes it is said (see,
e.g., p. 810 of Ref. \cite{Z-J}) that the global symmetry associated with the gauge group
is spontaneously broken. The precise meaning of this statement is, however, unclear to me.

\subsubsection*{Gauge invariant effective potential}

From lattice simulations it is known that the expectation value of $|\phi|^2$ is suited to
characterize the Higgs phase. It is, therefore, natural to couple the source in the
partition sum to this quantity, and consider the effective potential belonging to
\begin{equation}
Z_\Lambda=\int e^{-S_\Lambda + J\sum_{x\in\Lambda}|\phi_x|^2}~d\mu_\Lambda,
\end{equation}
where $d\mu_\Lambda$ is now the product of the normalized Haar measures and the Lebesque
measure $[d\phi]_\Lambda$. We emphasize that this effective potential should be regarded as
a function of $|\Phi|^2$, since this is the conjugate variable, and only as a function of
this variable it has to be convex. (This remark is, I believe, also relevant for the
Ginzburg-Landau free energy density of superconductivity, which is only convex as a
function of the \textit{square} of the absolute magnitude of the order parameter. Drawing
this free energy as a `Mexican hat' of the complex order parameter is somewhat misleading.)

For gauge models with the sum of the actions (40) and (46) the effective potential can be
computed numerically or in perturbation theory. The latter has been developed on the
lattice in close analogy to the continuum gauge theory \cite{MM}. We shall elaborate on
this in Sect. 3.1.

\section*{Appendices to Chapter 2}

\subsection*{Appendix A. The Legendre-Fenchel transform}

In thermodynamics and statistical mechanics it is important to generalize the usual
Legendre transformation to functions which are \textit{not everywhere differentiable}. This
generalization in a finite number of dimensions was introduced and developed by Fenchel in
1948.

The \textit{Legendre-Fenchel transform} $f^*$ of a real valued-function $f$ on $R^n$ is
defined as
\begin{equation}
f^{*}(x^{*})=\sup_{x\in R^n}\{\langle x,x^{*}\rangle -f(x)\},~~~x^{*}\in R^n.
\end{equation}
To be precise we have to assume that $f$ majorizes at least one convex function. (This
definition has an immediate generalization to dual pairs of \textit{infinite} dimensional
spaces.) The \textit{convex hull} of $f$ is then defined as the largest convex function
majorized by $f$ and is denoted by $cl(conv~f)$. In this situation $f^{*}$ is
\textit{convex} and the following statements hold:
\begin{equation}
f^{*}=(cl(conv~f))^{*},~~~f^{**}=cl(conv~f).
\end{equation}

For a proof and further information we  refer to \cite{Roc}. For a useful treatment, see
also \cite{Ell} or \cite{Si}.

If $f$ is convex and differentiable, this generalized Legendre transformation reduces to
the standard one (show this). In one dimension there is a simple geometrical construction
of $f^{*}$ (exercise).

\subsection*{Appendix B. Computation of a functional\\ determinant}

In this Appendix we compute the functional determinant of the operator
\begin{equation}
A=-\Delta +m^2 +P''(\Phi)
\end{equation}
that appeared in (23). It may be instructive to start from the lattice regularization and
then perform a (formal) continuum limit. The discrete Laplacian is naturally defined by
\begin{equation}
-\Delta=\sum_\alpha \partial_\alpha^\dag\partial_\alpha,
\end{equation}
where
\begin{equation}
(\partial_\alpha f)(x)= \frac{1}{a}[f(x+a e_\alpha)-f(x)], ~~~(\partial_\alpha^ \dag
f)(x)=\frac{1}{a}[f(x-a e_\alpha)-f(x)].
\end{equation}
Here $e_\alpha$ are the unit vectors in the lattice directions, and $a$ is the lattice
constant. The Laplacian becomes diagonal in the Fourier representation. On a finite
sublattice with $N_\alpha$ lattice points in the direction $e_\alpha $, the Fourier
transform of a function $f(x)$ is normalized as
\begin{equation}
\hat{f}(k)=\sum_{x\in\Lambda}f(x)~e^{-ik\cdot x}.
\end{equation}
The inverse transformation is given by the following sum over the discrete first Brillouin
zone $\Delta=\{k:~k_\alpha=(\pi/a)(2n_\alpha/N_\alpha),~-N_\alpha/2<n_\alpha\leq
N_\alpha/2\}$:
\begin{equation}
f(x)=\frac{1}{\left|\Lambda\right|}\sum_{k\in\Delta}\hat{f}(k)~e^{ik\cdot x}.
\end{equation}

One easily finds for the Laplacian in Fourier space:
\begin{equation}
-\Delta f \rightarrow ~\sum_\alpha\frac{2(1-\cos k_\alpha a)}{a^2}\hat{f}(k).
\end{equation}
The Fourier representation of the operator $A$ is therefore in $D$ dimensions
\begin{equation}
\hat{A}= diag~ \Bigl[\frac{2}{a^2}\sum_{\alpha=1}^{D} (1-\cos k_\alpha
a)+m^2+P''(\Phi)\Bigr ].
\end{equation}
Hence,
\begin{equation}
\ln\det A=\sum_{k\in\Delta} \ln\Bigl[ M^2+\frac{2}{a^2}\sum_{\alpha} (1-\cos k_{\alpha}
a)\Bigr],
\end{equation}
where $M^2=m^2+P''(\Phi)$. In the thermodynamic limit $N_1=N_2=N_3=:N\rightarrow \infty$,
we obtain, using the rule
\[\sum_{k\in\Delta}\rightarrow N^3(\frac{a}{2\pi})^3\int_{BZ}d^3k \sum_{k_4}~,  \]
with $BZ=[-\frac{\pi}{a},\frac{\pi}{a}]^3$:
\begin{equation}
\frac{\ln\det A}{(Na)^3}=\int_{BZ}\frac{d^3k}{(2\pi)^3}\sum_{k_4}\ln\Bigl[
M^2+\frac{2}{a^2}\sum_{\alpha} (1-\cos k_{\alpha} a)\Bigr].
\end{equation}
Here, $k_4=(2\pi/\beta) n, -\beta/2a< n \leq \beta/2a $.

This quantity diverges, of course, in the continuum limit $a\rightarrow 0$. We now show
that the temperature dependent part of the regularized free energy density
\[f_{reg}=\frac{1}{\beta(Na)^3}\frac{1}{2}~\ln\det\Bigl\{\frac{a^4}{2\pi}A\Bigr\}\]
converges to (24). Note that this formula reduces for $P=0$ to the correct free energy
density of a free scalar field with mass $m$.

For $a\rightarrow0$ we can replace the argument of the logarithm in (64) by
$\Omega^2(k)+k_4^2$. Thus
\begin{equation}
f_{reg}=\frac{1}{2\beta}\int_{BZ}\frac{d^3k}{(2\pi)^3}\sum_{n=-\beta/2a}^{\beta/2a}
\ln\Bigl\{\frac{a^4}{2\pi}\Bigl[\Omega^2(k)+\frac{4\pi^2n^2}{\beta^2}\Bigr]\Bigr\}.
\end{equation}
From this we subtract its value for $\beta\rightarrow \infty$. The sum appearing in (35)
can be worked out with the $\zeta$-function technique. We use
\begin{equation}
\sum_{n=-\beta/2a}^{\beta/2a}
\ln\Bigl[\Omega^2(k)+\frac{4\pi^2n^2}{\beta^2}\Bigr]=-\frac{d}{ds}\left.\zeta^{reg}(s)\right|_{s=0},
\end{equation}
where
\begin{equation}
\zeta^{reg}(s)=\sum_{n=-\beta/2a}^{\beta/2a}\frac{1}{\Bigl[\Omega^2(k)+\frac{4\pi^2n^2}{\beta^2}\Bigr]^s}.
\end{equation}
For $a\rightarrow 0, ~\zeta^{reg}$ converges to
\begin{equation}
\zeta(s)=\sum_{n=-\infty}^{\infty}\frac{1}{\Bigl[\Omega^2(k)+\frac{4\pi^2n^2}{\beta^2}\Bigr]^s}.
\end{equation}
This is the $\zeta$-function of the 1-dimensional differential operator
$-(d/d\tau)^2+\Omega^2$ on the interval $[0,\beta]$ with periodic boundary conditions.
Subtracting from $\zeta^{reg}(s)$ its value for $\beta \rightarrow\infty$ amounts for its
derivative at $s=0$ to the subtraction of a term linear in $\beta$, and thus of a
$\beta$-independent term for $f^{reg}$. Hence we need
\[ \frac{1}{\beta}\lim_{a\rightarrow
0}\Bigl[\frac{d\zeta^{reg}}{ds}-(\beta\rightarrow\infty)\Bigr]=\frac{d}{ds}\Bigl[\Bigl(\zeta(s)/\beta\Bigr)-
(\beta\rightarrow\infty)\Bigr] \] for $s=0$.

In the footnote below we show that
\begin{equation}
\zeta'(0)=-\beta\Omega-2\ln\Bigl(1-e^{-\beta\Omega}\Bigr).
\end{equation}
This proves our claim. Note that the first term in the last equation gives the zero-point
energy.

\subsubsection*{Footnote: Calculation of the $\zeta$-function}

We recall that the $\zeta$-function of a self-adjoint positive operator $Q$ with a pure
point spectrum and the spectral representation
\[ Q=\sum_{n} \lambda_n P_n, \]
where $P_n$ are the projections on the eigenspaces, is defined as \begin{equation}
\zeta_Q(s)=\sum_{s}\frac{g_n}{\lambda_n^s},~~~g_n=dim~P_n.
\end{equation}
From $\det Q=\exp(Tr\ln Q)$, we get
\begin{equation}
\ln\det Q=\sum_{n}g_n\ln\lambda_n=-\left.\frac{d\zeta_Q}{ds}\right|_{s=0}.
\end{equation}
Note also
\begin{equation}
\zeta_Q(s)=\sum_{n}\frac{1}{\Gamma(s)}\int_{0}^{\infty}dt~ t^{s-1}e^{-t\lambda_n}g_n
=\frac{1}{\Gamma(s)}\int_{0}^{\infty}dt~ t^{s-1}Tr\Bigl(e^{-tQ}\Bigr).
\end{equation}
For the operator $Q=-(d/d\tau)^2+\omega^2$ on the interval $[0,\beta]$ with periodic
boundary conditions, we obtain
\begin{eqnarray*}
\zeta_Q(s)&=& \beta^{2s}\sum_{n=-\infty}^{\infty}\frac{1}{\Bigl[\beta^2\omega^2+4\pi^2n^2\Bigr]^s}\\
& = &
\frac{\beta^{2s}}{\Gamma(s)}\int_{0}^{\infty}d\sigma\sigma^{s-1}\sum_{n=-\infty}^{\infty}
\exp\left[-\sigma(\beta^2\omega^2+4\pi^2n^2)\right] \\
& = &\omega^{-2s}+
\frac{2\beta^{2s}}{\Gamma(s)}\sum_{l=0}^{\infty}\frac{(\beta\omega)^{2l}}{l!}(-1)^l
\sum_{n=1}^{\infty}\int_{0}^{\infty}d\sigma~\sigma^{s+l-1}e^{-4\pi^2n^2\sigma}\\
& = & \omega^{-2s}+\frac{2\beta^{2s}}{(4\pi^2)^{s}}\zeta_R(2s) +
\frac{2\beta^{2s}}{\Gamma(s)}\sum_{l=1}^{\infty}\frac{(\beta\omega)^{2l}}{l!}\frac{(-1)^l}{(4\pi^2)^{s+l}}
\cdot\Gamma(s+l)\zeta_R(2s+2l),
\end{eqnarray*}
where $\zeta_R(s)$ is the Riemann $\zeta$-function
\[\zeta_R(s)=\sum_{n=1}^{\infty}\frac{1}{n^s}.\]
Since $\zeta_R(0)=-\frac{1}{2},~\zeta_R'(0)=-\frac{1}{2}\ln(2\pi)$ we have
\[ \zeta_Q'(0)=-2\ln(\beta\omega) +
2\sum_{l=1}^{\infty}\frac{(\beta\omega)^{2l}}{l}\frac{(-1)^l}{(4\pi^2)^{l}}~\zeta_R(2l).\]
The last factor can be expressed in terms of the Bernoulli numbers $B_{2l}$:
\begin{equation}
\zeta_R(2l)=(-1)^{l+1}\frac{(2\pi)^{2l}}{2(2l)!}B_{2l}.
\end{equation}
Therefore,
\[\zeta'_Q(0)=-2\ln(\beta\omega)-\sum_{l=1}^{\infty}\frac{(\beta\omega)^{2l}}{(2l)!}\frac{1}{l}B_{2l}.\]
Here we use the well-known relation
\[\coth(x)=\frac{1}{x}+2\sum_{l=1}^{\infty}\frac{(2x)^{2l-1}}{(2l)!}B_{2l}.\]
Integrating leads to
\[\int\coth(x)~dx=\ln x+\frac{1}{2}\sum_{l=1}^{\infty}\frac{(2x)^{2l}}{l(2l)!}B_{2l}.\]
Using this we obtain indeed (39):
\begin{equation}
\zeta_Q'(0)=-2\ln(\beta\omega)+2\ln\frac{\beta\omega}{2}-2\ln\sinh\frac{\beta\omega}{2}
=-\beta\omega -2\ln\Bigl(1-e^{-\beta\omega}\Bigr).
\end{equation}

\subsection*{Appendix C. General proof of Elitzur's\\ theorem}

It is quite easy to show generally that in a lattice gauge theory the expectation value of
any function, which depends only on the basic variables in a finite region (local
observable) and is non invariant under the gauge group, has a vanishing expectation value,
when a source term is set to zero \textit{after} the thermodynamic limit is taken.

In what follows, the basic dynamical fields (gauge fields $g_{x,y}$, Higgs fields $\phi_x$,
etc.) will collectively be denoted by $\varphi$. As in (44), we subtract from the gauge
invariant action $S[\varphi]$ a coupling $J\cdot\varphi$ to a (small) external source field
$J$. The expectation value of an `observable' $A[\varphi]$ in a region $\Lambda$ is, as
always,
\begin{equation}
\langle A\rangle_{\Lambda,J}=\frac{1}{Z_{\Lambda,J}}\int
\exp(-S_{\Lambda}+J\cdot\varphi)A~d\mu_\Lambda,~~~Z_{\Lambda,J}=\int
e^{-S_{\Lambda}+J\cdot\varphi}~d\mu_\Lambda,
\end{equation}
where $d\mu_\Lambda$ is the standard gauge invariant measure (involving the Haar measure
for $g_{x,y}$, etc.). We denote the action of a gauge transformation $g$ of the basic
fields by $\varphi\mapsto T(g)\varphi$. Non-invariance of a local $A[\varphi]$, depending
only on field configurations belonging to a finite subset of the lattice, means that
\begin{equation}
\int A[T(g)\varphi]~d\mu_\Lambda=0,
\end{equation}
because the integral on the left gives the invariant component of $A[\varphi]$.

We split the field configurations into two complementary sets $\{\varphi'\}$ and
$\{\varphi''\}$, where the first set denotes those on which $A$ depends. Associated to this
we consider the subgroup of gauge transformations which keeps the elements of the second
set fixed. If $(dg)_\Lambda$ denotes the product of the Haar measures corresponding to this
subgroup, we have
\[\langle A\rangle_{\Lambda,J}=\frac{1}{Z_{\Lambda,J}}\int
\exp(-S_{\Lambda}[\varphi]+J'\cdot
T(g)\varphi'+J''\cdot\varphi'')A[T(g)\varphi']~d\mu_\Lambda(\varphi) (dg)_\Lambda. \]
Because of (72) we can subtract 1 from the factor $\exp[J'\cdot T(g)\varphi']$ in the
integrand, hence the integral vanishes for $J\rightarrow 0$, for `reasonable' observables
$A$, uniformly in $\Lambda$ and $\beta$. Since the partition sum remains uniformly positive
in this limit, we arrive at
\begin{equation}
\langle A\rangle=\lim_{J\rightarrow 0}\lim_{\Lambda\rightarrow Z^D}\langle
A\rangle_{\Lambda,J}=0.
\end{equation}

Intuitively one can perhaps say, that a gauge theory cannot be spontaneously broken,
because local symmetry transformations involve only ``local changes'', and hence the
transformed states are not disjoint, similarly as for systems with a finite number of
degrees of freedom. Unfortunately, the commonly used terminology in most texts does not
describe the situation adequately.

\section{Applications to cosmological phase\\ transitions}

We now turn to relevant applications of the general theory to cosmological phase
transitions. I shall begin with a relatively detailed discussion of the electroweak phase
transition, on which much effort gas been invested in recent years.

\subsection{Thermodynamics of the electroweak phase transition}

One of the main motivations for the large amount of work was the suggestion that the
observed baryon asymmetry in the Universe was eventually determined at this transition.
Indeed, all three Sakharov conditions for baryogenesis are, in principle, fulfilled. One of
them, namely the departure from equilibrium, depends on the nature of the phase transition.

Out of equilibrium processes require a first order phase transition. We shall see that this
would be realized for Higgs masses much below the current experimental lower limit. With
increasing Higgs mass the phase transition weakens rapidly and presumably turns into a
smooth crossover. For realistic Higgs masses it becomes too weak for baryogenesis. In
supersymmetric extensions of the Standard Model there remains a small parameter range for
which the transition is sufficiently strong.

\subsubsection{The $SU(2)$-Higgs model}

Reliable results for the electroweak phase transition can only come from non-perturbative
methods, that is from lattice simulations, because the Higgs field is nearly massless at
the transition which leads to serious infrared divergencies in the perturbation expansion.
A complete lattice simulation of the Standard Model is not feasible due to the presence of
chiral fermions. The main (infrared) problems are, however, only connected to the bosonic
sector. For this reason people usually studied simulations of the $SU(2)$-Higgs model and
used perturbation calculations to include the $U(1)$ gauge group and fermions.

In what follows, I shall describe some results of such studies. For practical reasons
(avoidance of huge lattices), people used in recent years asymmetric lattices (different
spacings in temporal and spatial directions; see e.g. \cite{Fo}). For the sake of
presentation, I shall ignore the complications connected with this.

\subsubsection*{The lattice action}

In writing down the action we adapt the previously used notation to what has become
standard.\footnote{Slight remaining differences to some of the quoted papers amount to the
replacement $U_{x\mu}\rightarrow U_{x\mu}^\dag=U_{x\mu}^{-1}$.} The bond variables of the
gauge group $SU(2)$ are now denoted by $U_{x\mu}$ for the bond $\langle
x,x+\hat{\mu}\rangle$, where $\hat{\mu}$ stands for unit vector $e_\mu$. The Higgs doublet
can be described by a $2\times 2$ complex matrix $\varphi_x$, satisfying the
pseudo-hermiticity condition
\[ \varphi_x^\dagger=\varepsilon\varphi_x^T\varepsilon^{-1},~~~\varepsilon=\left(\begin{array}{cc}
0 & 1\\-1 & 0 \end{array} \right) .\] Under a gauge transformation $x\mapsto V_x$ the Higgs
field transforms as $\varphi_x\mapsto V_x\varphi_x$. Equivalently, the Higgs doublet ca be
described by four real scalar fields $\phi_S$, related to $\varphi$ by
\[ \varphi=\phi_0 1_2 + i\phi_k\tau_k ,\]
where the $\tau_k$ are the Pauli matrices.

In the formulae below we set the lattice spacing $a$ equal to unity. The action is
parametrized as follows:
\begin{eqnarray}
\lefteqn{S[U,\varphi]=\beta\sum_{P}(1-\frac{1}{2}Tr~U_P)}\nonumber \\
 & & + \sum_{x}\Bigl\{\frac{1}{2}Tr(\varphi_x^\dagger\varphi_x)+\lambda\Bigl[\frac{1}{2}
Tr(\varphi_x^\dagger\varphi_x)-1\Bigr]^2-\kappa\sum_{\mu=1}^4 Tr(\varphi_{x}^\dagger
U_{x\mu}\varphi_{x+\hat{\mu}}) \Bigr\}
\end{eqnarray}
Here, the parameter $\beta$ (not to be confused with $1/kT$) is related to the gauge
coupling constant $g$ by $\beta=4/g^2$. The last term is the hopping part of the gauge
invariant kinetic contribution of the scalar field. The hopping parameter $\kappa$ is
related to the bare scalar mass $m_0$ by $m_0^2=(1-8\kappa-2\lambda)/\kappa $ (show this).
As mentioned earlier, the gauge invariant length of the Higgs field,
$\rho_x^2:=\frac{1}{2}Tr(\varphi_x^\dagger\varphi_x)$, plays an important role. It turns
out that this quantity has large autocorrelations and can be used to characterize the Higgs
phase (as in the standard semiclassical treatment of the Higgs mechanism).

\subsubsection*{Gauge invariant effective potential}

An important and useful object for investigating phase transitions is the gauge invariant
effective potential (53). For the present model we have to start from the partition sum
\begin{equation}
Z_\Lambda=\int e^{-S_\Lambda + J\sum_{x\in\Lambda}\rho_x^2}~d\mu_\Lambda,
\end{equation}
with the action (78) which we rewrite, using the polar decomposition
\begin{equation}
\varphi_x=\rho_x\alpha_x,~~~\alpha_x\in SU(2),
\end{equation}
as
\begin{eqnarray}
\lefteqn{S[U,\varphi]=\beta\sum_{P}(1-\frac{1}{2}Tr~U_P)}\nonumber \\
 & & + \sum_{x}\Bigl\{(\rho_x)^2+\lambda(\rho_x^2-1)^2
-\kappa\sum_{\mu=1}^4\rho_{x}\rho_{x+\hat{\mu}} Tr(\alpha_{x})^\dagger
U_{x\mu}\alpha_{x+\hat{\mu}}) \Bigr\}.
\end{eqnarray}
The expression in the last trace is a gauge invariant link variable, which we call
$V_{x\mu}$. Because of the gauge invariance of the gauge action $S_g[U]$, we find for the
total action
\begin{equation}
S[V,\rho]= S_g[V]+\sum_{x}\Bigl\{\rho_x^2+\lambda(\rho_x^2-1)^2 -\kappa\sum_{\mu=1}^4
\rho_{x}\rho_{x+\hat{\mu}} Tr(V_{x\mu}) \Bigr\}.
\end{equation}

Next, we have to discuss the integration measure. From the link variables $V_{x\mu}$ we get
the product of the Haar measures $dV_{x\mu}$. For the Higgs field we have from every
lattice point the product of the Lebesque measures for the real fields,
\[ d\varphi_x=\prod_{S} d\phi_S(x)=\rho_x^3 d\rho_x d\sigma_{\phi_x},\]
where $d\sigma_\phi$ is the natural measure of the three sphere in $\phi$-space. Using (80)
this can be written as
\[d\varphi_x=\rho_x^3 d\rho_x 2\pi^2 d\alpha_x,\]
where $d\alpha_x$ is again the Haar measure. Since the action does not depend on $\alpha_x$
we can trivially integrate over the $d\alpha_x$ and obtain for $d\mu$ in (79), up to an
irrelevant normalization,
\begin{equation}
d\mu = \prod_{x,\mu}dV_{x\mu}\prod_{x}\rho_x^3 d\rho_x.
\end{equation}
Eqs. (79), (82) and (83) determine the thermodynamics of the $SU(2)$ Higgs model. Note that
-- without any gauge fixing -- three scalars have disappeared. (Repeat this in the
continuum theory.) The action (82) and the measure (83) have a global $SU(2)$ invariance:
$\rho_x\mapsto \rho_x,~V_{x\mu}\mapsto UV_{x\mu}U^{-1},~U\in SU(2)$ (\textit{weak
isospin}). This symmetry is reflected in the mass spectrum.

The phase structure of the model will be discussed shortly. Let us consider a first order
phase transition between the `symmetric' (s) and the `broken' (b) phases. Along the
coexistence curve $J(T)$ the free energy densities are the same:
\[ w_s(T,J(T))=w_b(T,J(T)). \]
Differentiating with respect to $T$ gives (note that $w(T,J)$ is the negative of the
Helmholtz free energy density) the following \textit{Clausius-Clapeyron equation} between
the latent heat $\Delta Q$ and the jump $\Delta\rho^2$ in the order parameter
\begin{equation}
\Delta Q=\Delta\rho^2 T\frac{dJ}{dT}.
\end{equation}

\subsubsection*{Phase structure}

Numerical studies established for a fixed $\lambda=O(1)$ the following picture (see Fig. 4)
of the phase diagram in the 3-dimensional parameter space $\beta,~\kappa,~T$, where $T$ is
the temperature in lattice units (inverse of the time extent of the lattice):

For $\kappa=0$ the model reduces to a pure $SU(2)$ gauge theory. As indicated in Fig. 4,
there is for this limiting model for any $T>0$ a deconfinement phase transition at a
critical value $\beta_c$ which increases with decreasing temperature. This phase transition
also exists when $\kappa$ is switched on, but for large values of the hopping parameter the
transition may change to an analytical crossover. In the limit $\beta=\infty$ the gauge
fields become trivial (pure gauge). Setting them to unity, we get a $\phi^4$ theory for an
$SU(2)$ Higgs field, having a global $SU(2)_L\times SU(2)_R$ symmetry. For a fixed positive
temperature this model exhibits a phase transition for a critical value $\kappa_c$, above
which the symmetry is broken down to $O(3)$. Simulations showed always a second order phase
transition. With varying temperature a transition line is found. This develops into a
surface for finite values of $\beta$ where the electroweak phase transition takes place.
When $\beta$ becomes small, one expects a region in parameter space where no phase
transition occurs. So far $\lambda$ was kept fixed. When both $\lambda$ and the temperature
become large enough, the electroweak transition may turn into a crossover phenomenon.

\begin{figure}
\begin{center}
\includegraphics[height=0.4\textheight]{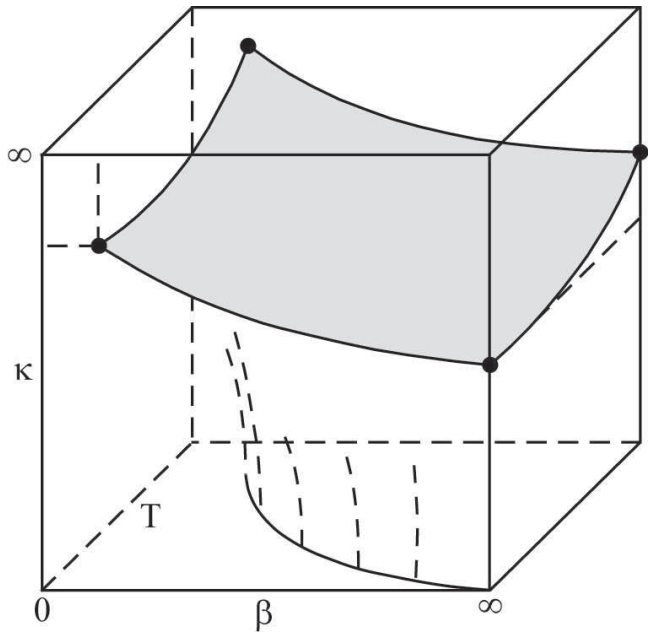}
\caption{Phase diagram for the $SU(2)$ Higgs model. For a fixed quartic coupling
$\lambda=O(1)$ there are three regions in the $\beta,\kappa,T$ parameter space:
\textit{confinement region} for $\beta<\beta_c,~\kappa\ll1$; \textit{Higgs region} for
$\beta\gg1,~\kappa>\kappa_c$; \textit{deconfinement region} for
$\beta>\beta_c,~\kappa<\kappa_c$.}
\label{Fig-4}
\end{center}
\end{figure}

For $T=0$ the phase structure is sketched in Fig. 5 in the parameter space
$\beta,\kappa,\lambda$. For small values of $\kappa$ the system is in the confinement phase
(at $\kappa=0$ the model reduces to a pure $SU(2)$ gauge theory). Along the surface
$\Sigma$ there is a first order phase transition to the Higgs phase (with one spin-0 Higgs
boson and three massive gauge bosons), except at the boundary $\beta=\infty~ (g=0)$. The
boundary $\partial\Sigma_1$ of is a second order transition line as for $T>0$. For small
$\beta$ (large coupling) and relatively large values of the hopping parameter $\kappa$
there is an analytic connection between the two regions beyond the boundary piece
$\partial\Sigma_2$. (This has been established rigorously in \cite{OSl}.) The distinction
between confinement and Higgs phases has then no  meaning anymore.

If one uses the actual value of the weak coupling constant, it turns out that the finite
temperature phase transition always takes place between the Higgs and the deconfinement
regions.

\begin{figure}
\begin{center}
\includegraphics[height=0.4\textheight]{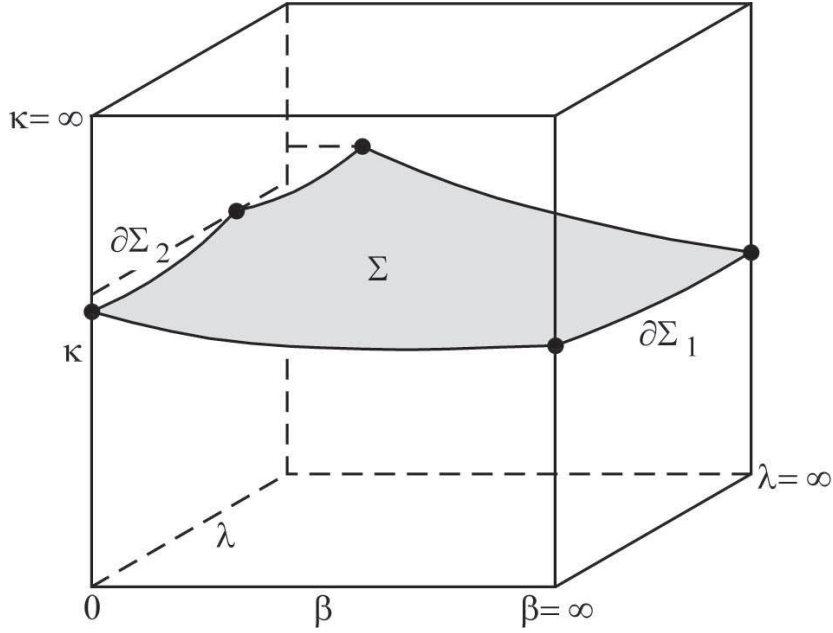}
\caption{Phase structure for zero temperature (adapted from \cite{MM}, Sect. 6.1).}
\label{Fig-5}
\end{center}
\end{figure}

\subsubsection*{One-loop lattice calculation of the effective potential}

The effective gauge invariant potential of the $SU(2)$ Higgs model has been determined
numerically, along with other quantities \cite{Mo}. Here, we want to derive the one-loop
approximation of this important object. This is an instructive example of \textit{lattice
perturbation theory}.

Let us first consider the `broken' phase. The exponent in (79) has for $V_{x\mu}=1$ a
stationary point for the value $\bar{\rho}$ given by
\begin{equation}
\bar{\rho}^2=1+\frac{8\kappa+J-1}{2\lambda},
\end{equation}
provided $J>1-8\kappa-2\lambda$. In the tree approximation (Laplace approximation of the
infinite volume integral) the effective potential is
\begin{equation}
V_{tree}(\Phi^2)= (1-8\kappa)\Phi^2 +\lambda(\Phi^2-1)^2.
\end{equation}
(This is convex as a function of $\Phi^2$ !) For the one-loop correction we set
$\rho_x=\bar{\rho}+\sigma_x,~U_{x\mu}=\exp(A_\mu(x))$ and consider $A_\mu(x)$ as small. The
quadratic piece in $\sigma_x$ of the exponent in (79) is just the action of a free scalar
lattice field $\sigma$ with effective mass
\begin{equation}
m_\varphi^2=\frac{4}{\kappa}\lambda\bar{\rho}^2=m_{H,0}\bar{\rho}^2/v^2,
\end{equation}
where $v$ is the value of $|\Phi|$ at the minimum of $V_{tree}(\Phi^2)$, and $m_{H,0}$ is
the zeroth order Higgs mass. According to Sect. 2.2 this gives the following contribution
to the one-loop correction
\begin{equation}
V_{1-loop}^\varphi(\Phi^2)=\int_{BZ}\frac{d^4k}{(2\pi)^4}\frac{1}{2}\ln(\hat{k}^2+
m_H^2\Phi^2/v^2),
\end{equation}
where $\hat{k}^2:=\sum_{\mu}2(1-\cos k_\mu)$. For $T>0$ the integral over $k_4$ has to be
replaced by $T$ times the sum over $k_4$, with $k_4=(2\pi T)n,~n\in \mathbf{Z}$.

For the contribution of the gauge field we set $V_{x\mu}=\exp(A_\mu(x))$ and consider
$A_\mu(x)$ to be small. Then the gauge part of the action reduces in quadratic
approximation to the Maxwell action of the three scalar fields $A_{\mu}^{a}$, defined by
$A_{\mu}=ig A_{\mu}^{a}\tau_{a}/2$. These receive, as in the continuum theory, a mass term
from the coupling to the scalar field (last term in (82)) with effective mass
\begin{equation}
m_g^2=\frac{1}{2}\kappa g^2\bar{\rho}^2=m_{W,0}^2\bar{\rho}^2/v^2,
\end{equation}
where $m_{W,0}$ is the zeroth order W-mass.

Next the measure (83) has to be rewritten in the appropriate approximation in terms of
$A_\mu(x)$ and $\sigma_x$. It is a nice exercise in group theory to express the normalized
Haar measure $dU$ of $SU(2)$ in terms of the canonical coordinates (of the first kind)
$X_a$, defined by $U=\exp(iX_a\tau_a/2)$. The result is
\begin{equation}
dU=\frac{1}{4\pi^2}\frac{\sin^2(|X|/2)}{|X|^2}d^3X.
\end{equation}
Hence,
\[dV_{x\mu}=\frac{1}{4\pi^2}\frac{\sin^2(g|A_\mu|/2)}{g^2|A_\mu|^2}d^3A_\mu^a~, \]
with $|A_\mu|^2=A_\mu^a A_\mu^a$. We write the first factor in (83) as
\[\prod_{x,\mu}dV_{x\mu}=e^{-S_m}\prod_{x,\mu}dA_\mu^a~,\]
with
\begin{equation}
S_m=-\sum_{x,\mu}\ln\left(\frac{1}{4\pi^2}\frac{\sin^2(g|A_\mu|/2)}{g^2|A_\mu|^2}\right)=
\sum_{x,\mu}\frac{g^2}{12}A_{\mu}^{a}A_\mu^{a}+O(g^4).
\end{equation}
(An additive numerical constant was dropped in the last expression.)

In sufficient approximation the measure (83) becomes, up to a normalization,
\begin{equation}
d\mu=\exp\left\{-\sum_{x,\mu,a}\frac{g^2}{12}A_{\mu}^{a}A_\mu^{a}\right\}
\prod_{x,\mu,a}dA_\mu^a(x)\prod_{x}d\sigma_x~.
\end{equation}
The exponential in this expression is a mass term with  mass $g^2/6$. This acts like a
mass-counter term and diverges in the limit $a\rightarrow0$.  We encounter here an example
of a general phenomenon: The lattice regularization provides its own counter terms which
ensure renormalizability of the theory.

The contribution of the free massive vector fields $A_\mu(x)$ with the effective mass
$m_g^2+g^2/6$ gives the following contribution to the one-loop potential (exercise)
\begin{equation}
V_{1-loop}^g(\Phi^2)=9\int_{BZ}\frac{d^4k}{(2\pi)^4}\frac{1}{2}\ln(\hat{k}^2+
m_W^2\Phi^2/v^2 +g^2/6).
\end{equation}
The reason for the factor 9 is obvious: three massive vector fields with three degrees of
freedom. The additive term $g^2/6$ in the argument of the logarithm would not appear in a
continuum calculation.

We consider now the `symmetric' phase. There is a stationary point at $\bar{\rho}=0$, for
which the tree level potential vanishes. The exponent in (79) is now (apart from a sign)
the action of a free scalar lattice field $\rho_x$ with effective $J$-dependent mass
$m_s^2=m_0^2 -J/\kappa$, plus that of three massless vector fields $A_\mu(x)$ which get an
effective mass $g^2/6$ from the link integration measure in (83). We have to keep in mind
that $\rho_x$ is a radial variable and that the measure (83) contains the weight factors
$\rho_x^3$. In the steepest decent approximation we obtain for the one-loop free energy
density, apart from $J$-independent contributions due to the gauge fields,
\begin{equation}
w(J)\simeq -\int_{BZ}\frac{d^4k}{(2\pi)^4}\frac{1}{2}\ln(\hat{k}^2+2m_s^2).
\end{equation}
The one-loop effective potential is the Legendre transform of $w(J)~ (\Phi^2$ is the
conjugate variable). This has to be done numerically.

At the critical temperature the minima of the two effective potentials are degenerate. The
critical temperature in units of the Higgs mass as a function of the Higgs mass is shown in
the upper part of Fig. 6, both for numerical simulations and perturbation
calculations.\footnote{We do not discuss here the so-called \textit{effective field theory}
method, which makes use of the compactness of the time interval for positive temperatures.
This suggests, in complete analogy to the dimensional reduction in Kaluza-Klein theories,
the construction of a $3D$ effective action containing the zero modes of bosonic fields.
For an introduction to this method ( which can be regarded as a generalization of the high
temperature expansion), as well as for extensive references, see \cite{Mi}.}

\begin{figure}
\begin{center}
\includegraphics[height=0.75\textheight]{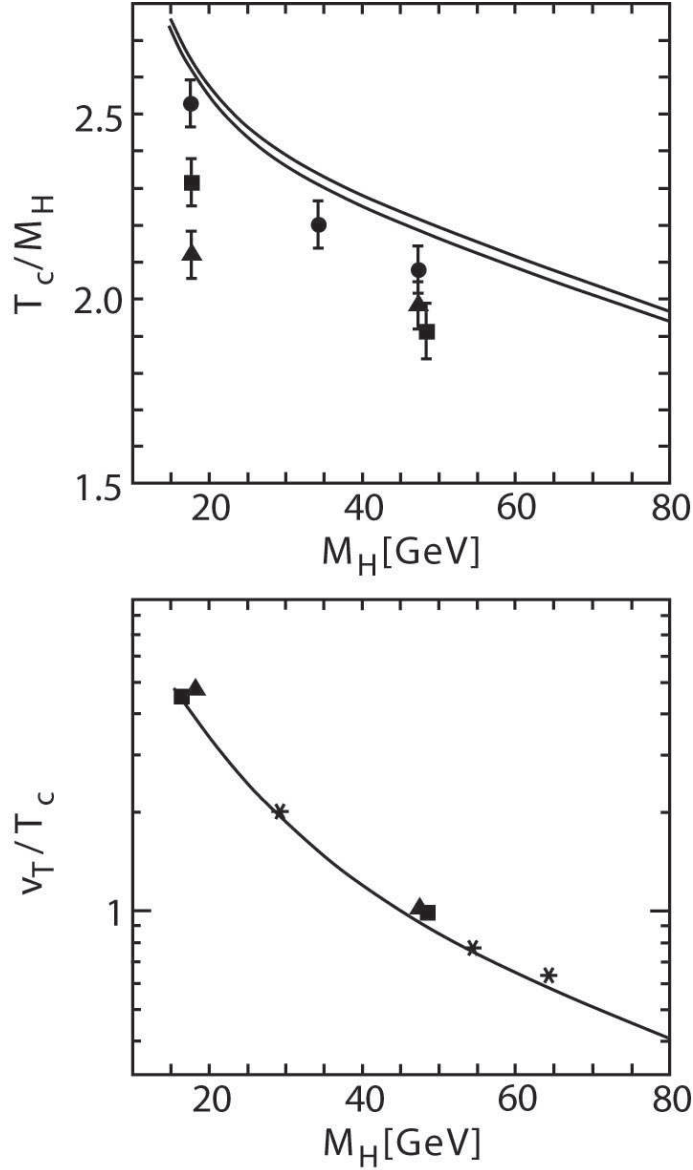}
\caption{(\textit{top panel}): Critical temperature as a function of the Higgs mass $m_H$.
The full squares and triangles are from 4-dimensional simulations with different lattice
extensions in the imaginary time direction. The full circles are results from lattice
perturbation theory, extrapolated to the continuum. The solid lines are from continuum
2-loop calculations \cite{Bu}. (From Ref. \cite{Ja}.) (\textit{bottom panel}): Jump of the
order parameter at the critical temperature. The triangles and squares are from the same
simulations as in the top panel, and the solid line is again the result of a continuum
2-loop calculation. The starred symbols are from simulations of the so-called 3-dimensional
`reduced model' (see footnote 4). } \label{Fig-6}
\end{center}
\end{figure}

The jump $\Delta$ in the gauge invariant quantity $\rho_x^2,~
v_T:=(\Delta\langle\rho_x^2\rangle)^{1/2}$, is a measure of the strength of the first order
phase transition. Its dependence on the Higgs mass is shown in the lower part of Fig. 6. A
similar plot for the latent heat is given in Fig. 7. More recently it has been shown
\cite{Fo} that the first order transition ends at $m_H=66.5\pm 1.4~GeV$. Above this
endpoint only a rapid cross-over can be seen. This number increases for the Standard Model
to $72.1\pm1.4~GeV$. In the framework of the effective field theory method (footnote 4)
this result has been obtained in \cite{Kaj}, \cite{Rum} and \cite{La1} (for a short review,
see \cite{La2}).

\begin{figure}
\begin{center}
\includegraphics[height=0.4\textheight]{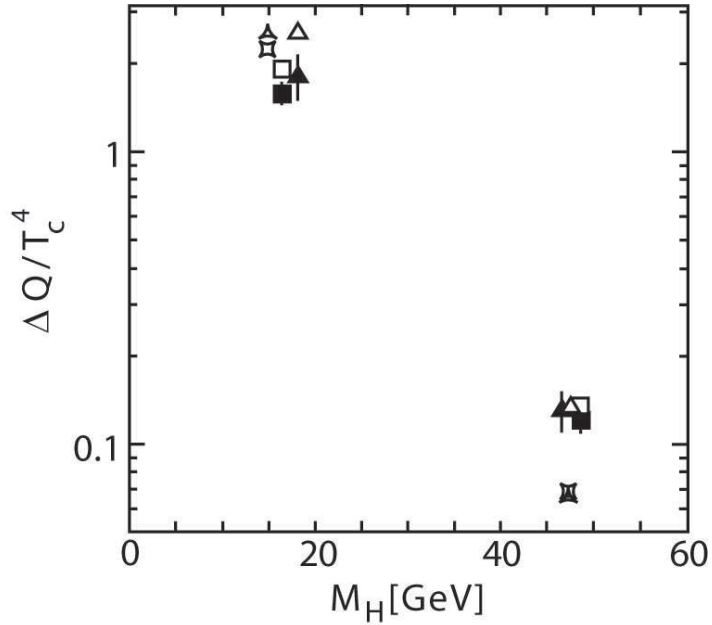}
\caption{Latent heat as a function of the Higgs mass. The symbols have the same meaning as
in the previous figures.} \label{Fig-7}
\end{center}
\end{figure}

\subsection{Baryogenesis and electroweak phase transition}

The previous results, in particular Fig.6, are crucial for answering the question whether
the cosmological baryon asymmetry might have been generated in the electroweak phase
transition. We shall see that this was not the case. Nevertheless, fermion-number violating
electroweak interactions are of interest, since they changed the baryon and lepton
asymmetries that were generated during earlier epochs, expected in the context of unified
extensions of electroweak and strong interactions (GUTs). For reviews, see \cite{Co} and
\cite{DKu}.

\subsubsection*{Anomalous fermion number violations}

Let me first recall why baryon number (B) and lepton number (L) are not conserved in the
Standard Model, although they are obviously conserved in perturbation theory. On the
non-perturbative level this due to the chiral nature of the weak interactions. Indeed, this
implies that the divergences of the baryon and lepton currents are both anomalous (quantum
mechanically non-conserved). It turns out that this violation is such that $B-L$ is
conserved.

Consider, for illustration, in the $SU(2)$ Higgs model a massless left-handed fermion
doublet $\psi_L$ and the global $U(1)$ fermion current
\[ J^\mu=\bar{\psi}_L\gamma^\mu\psi_L.\]
Classically, this is conserved, and expresses the conservation of the number of fermions
corresponding to the field $\psi_L$. In quantum theory, the famous triangle anomaly of
Adler, Bell and Jackiw leads to the modification\footnote{Thanks to the work of M.
L\"{u}scher \cite{L}, this can now also be derived on the lattice in complete analogy to
the Fujikawa method for the continuum theory. For review and references, see \cite{CW}.}
\begin{equation}
\partial_\mu J^\mu = -\frac{1}{16\pi^2}tr(F_{\mu\nu} \tilde{F}^{\mu\nu}),
\end{equation}
where $\tilde{F}_{\mu\nu}$ is the dual field strength of the $SU(2)$ gauge field
$F_{\mu\nu}$, belonging to the gauge potential $A_\mu(x)=-igA_\mu^a\tau_a/2$ (see, e.g.,
Chap. 19 of \cite{Pe} or \cite{Na} ). The right-hand side of this equation, the famous
characteristic second \textit{Chern density}, has an important topological meaning.  Its
integral (Chern number) is equal to a certain winding number, as shown below. Locally, it
is a divergence of a current. It is more elegant to write this in terms of differential
forms. The Chern 4-form belonging to the right-hand side of (95)
\begin{equation}
c_2(A)=\frac{1}{8\pi^2}tr(F\wedge F)~~~(F=\frac{1}{2}F_{\mu\nu}dx^\mu\wedge dx^\nu)
\end{equation}
is (modulo global questions) an exact differential
\begin{equation}
c_2=dQ(A),~~~Q(A)=\frac{1}{8\pi^2}Tr\Bigl[A\wedge(dA+\frac{2}{3}A\wedge A)\Bigr].
\end{equation}
$Q(A)$ is the \textit{Chern-Simons form} ($A=A_\mu dx^\mu$). This 3-form is not gauge
invariant. If we integrate (95), i.e.,
\begin{equation}
d\star J=c_2(A)
\end{equation}
over a region $\mathcal{D}$ of space-time between two time
slices, Stokes' theorem gives
\[ \int_{\mathcal{D}}d\star J=\int_{t=t_2}\star J -\int_{t=t_1}\star J=\Delta N_{CS},\]
where $N_{CS}$ is the difference of the Chern-Simons numbers
\begin{equation}
\Delta N_{CS}[A]=\int_{t=t_2}Q(A) -\int_{t=t_1}Q(A).
\end{equation}
This difference is gauge invariant since it is equal to the integral of $c_2(A)$ over
$\mathcal{D}$. We see that the change of the fermion number $\Delta N_F$ associated to
$\psi_L$ is equal to $\Delta N_{CS}$:
\begin{equation}
\Delta N_F=\Delta N_{CS}[A].
\end{equation}

Before further elaborating on this, we consider the left-handed doublets of the three
families of leptons and quarks of the Standard Model. The previous equation implies the
same fermion-number change for each species. For the leptons the changes of lepton number
are
\begin{equation}
\Delta N_e=\Delta N_\mu=\Delta N_\tau=\Delta N_{CS}[A],
\end{equation}
and the changes for the quarks imply the following change of the baryon number
\begin{equation}
\Delta B=\frac{1}{3}\cdot 3 \cdot 3~\Delta N_{CS}[A]
\end{equation}
($\frac{1}{3}$ from the baryon number of quarks, $3\cdot3$ from the three colors and number
of generations). This also shows that $B-L$ is conserved.

The non-perturbative effects we are discussing involve strong gauge fields, and it is
therefore reasonable to treat them classically. So let us now consider classical field
configurations and the changes of their Chern-Simons numbers in time. In particular, we are
interested in field configurations which interpolate between different classical vacua with
pure gauges $A_\mu=U^{-1}\partial_\mu U,~ U(x)$ an $SU(2)$-valued function. The
Chern-Simons number of such a vacuum gauge field is in general different from zero. A short
calculation gives
\begin{equation}
Q(A)=-\frac{1}{24\pi^2}Tr\Bigl(U^{-1}dU\wedge U^{-1}dU\wedge U^{-1}dU\Bigr).
\end{equation}
We assume that for fixed time $U(x)$ converges at infinity to $1_2$ in such a manner that
we can regard it as a function on the compactified space $S^3$. Since $SU(2)$ is
topologically also a three-sphere, we then have a map from $S^3$ to $S^3$. We now show that
the winding number (degree) of this map is just the Chern-Simons number of the pure gauge.
The easiest way to see this is to note first that $U^{-1}dU$ is the pull-back by $U$ of the
Maurer-Cartan form $\theta$ on $SU(2)$:
\[U^{-1}dU=U^\star(\theta).\]
Therefore, Eq.(103) can be written as
\begin{equation}
Q(U^{-1}dU)=-\frac{1}{24\pi^2}U^\star\Bigl[ Tr(\theta\wedge\theta\wedge\theta)\Bigr].
\end{equation}
A general result on winding numbers tells us that the one for $U:~S^3\rightarrow S^3$ is
\[n(U)=\frac{\int_{S^3}U^\star(\omega)}{\int_{SU(2)}(\omega)}\]
for any volume form $\omega$ on $S^3$ (see, e.g., Sect. 7.5 of \cite{Ma}), in particular
for $\omega=Tr(\theta\wedge\theta\wedge\theta)$. Hence,
\[N_{CS}[U^{-1}dU]= n(U)(-\frac{1}{24\pi^2})\int_{SU(2)}\omega .\]
But $(-1/24\pi^2)\omega$ is the three-form corresponding to the normalized Haar measure,
whence
\begin{equation}
N_{CS}[U^{-1}dU]=n(U).
\end{equation}
This integer characterizes topologically different vacua.

As a result of these considerations we obtain for a field configuration which interpolates
between vacua at $t=-\infty$ and $t=+\infty$
\begin{equation}
\Delta N_{CS}[U^{-1}dU]=\left.n(U)\right|_{t=\infty}-\left.n(U)\right|_{t=-\infty}.
\end{equation}
This difference is also equal to the second Chern number
\[ c_2=\int c_2(A),\]
which is, of course, gauge invariant.

The winding number $n(U)$ also shows up in the vacuum configuration of the Higgs field. It
is straightforward to verify the following identity (we omit the wedge symbols):
\begin{equation}
Tr[(\phi^\dagger d\phi)^3]-3Tr[A(dA+\frac{2}{3}A^2)]=Tr[(\phi^\dagger
D\phi)^3]-3Tr(\phi\dagger F D\phi)+3 d(A~ d\phi~\phi^\dagger).
\end{equation}
This implies that the left-hand side is gauge invariant, up to an exact differential. Apart
from this exact differential the right-hand side vanishes for a classical vacuum, where
$D\phi=0$. Integrating the last equation we see that the winding number of the Higgs vacuum
field \begin{equation} N_H=\frac{1}{24\pi^2}\int Tr[(\phi^\dagger d\phi)^3]
\end{equation}
is equal to the Chern-Simons number, and for a vacuum-vacuum transition we obtain the gauge
invariant relation
\begin{equation}
\Delta N_{CS}=\Delta N_H .
\end{equation}

For the smallest non-vanishing change (102) gives
\begin{equation}
\Delta B=-\Delta L=3,
\end{equation}
the quantum numbers of 12 left-handed fermions. States such as $\prod u_Ld_Ld_L\nu_L$,
where $\Pi$ denotes the product over all generations, can be created out of the vacuum.
(Altogether there are 12 such combinations.)

Vacua with different topological numbers $n$ are separated by a potential barrier, because
a transition has to involve non-vacuum fields. At zero temperature and `low' energies such
a transition is a tunnelling process that can be estimated by using (constraint)
instantons. For a pure Yang-Mills theory an instanton minimizes the Euclidean action in the
sector with Chern number $c_2=1$. For $SU(2)$ the corresponding action is equal to
$8\pi^2/g^2$, hence the tunnelling probability is enormously suppressed by the factor
$\exp(-4\pi/\alpha_W)\sim 10^{-170}$, where
$\alpha_W=g^2/4\pi=\alpha/\sin^2\theta_W\simeq1/29~(\theta_W$ is the weak mixing angle).
When the system is in the Higgs phase things are somewhat more complicated, since there is
then no minimum for the Euclidean action with $c_2=1$. Nevertheless, the tunnelling rate is
suppressed by the same exponential factor. (For details, see \cite{Af}.)

Things are very different at high temperatures, corresponding to energy scales set by the
height of the barrier. This is given by the static saddle point solution of the bosonic
sector (Yang-Mills-Higgs equations), the so-called \textit{sphaleron solution}. The energy
of this spherically symmetric unstable solution is \cite{Kl}
\begin{equation}
E_{sph}=\frac{2m_W}{\alpha_W} f(\frac{m_H}{m_W}),
\end{equation}
where the function varies from 1.56 to 2.72 as $m_H/m_W$ varies from zero to large values.
The height of the barrier is thus of order 10 TeV. At temperatures larger than the critical
temperature $T_c\sim m_H/g,  B$-violating processes may proceed through thermal activation
over the barrier which is then much lower because the $W$ boson mass eventually vanishes.
The tunnelling suppression factor $\exp(-E_{sph}/T)$ is no more effective in the symmetric
phase. Detailed studies \cite{ASY} have established that the naive power counting estimate
for the rate per unit volume in the unbroken phase
\begin{equation}
\Gamma= \mathcal{O}(1)(\alpha_W T)^4
\end{equation}
indeed holds. (For a review, see \cite{Sh}.) We compare this with the expansion rate $H$ of
the Universe. During these early phases the energy density $\rho$ is dominated by
relativistic particles, thus $\rho=g_\ast(\pi^2/30) T^4$, where $g_\ast$ is the effective
number of degrees of freedom per particle (2 for the photon). The Friedmann equation gives
\[ H\simeq g_{\ast}^{1/2}T^2/M_{Planck} \]
with $g_{\ast}^{1/2}=\mathcal{O}(10)$. For dimensional reasons, $\Gamma$ should be compared
with the time derivative of $T^3$ ( a typical particle density), i.e. with $HT^3$. (We used
that $T$ is most of the time inversely proportional to the expansion factor.) The ratio
\[ \frac{\Gamma}{H T^3} \sim  \alpha_{W}^4\frac{M_{Planck}}{T} \]
is much larger than 1 for $T_c<T<10^{12}~GeV$. In the symmetric phase the $B$-violating
reactions are thus fast enough to maintain thermal equilibrium.

During this phase the $B+L$ part of $B=\frac{1}{2}(B-L)+\frac{1}{2}(B+L)$ is therefore
changed. At first sight one might expect that it is completely washed out, and that
$B=-L=\frac{1}{2}(B-L)_{initial}$ is established. Things are, however, somewhat more
complicated, since only the fermion currents of the left-handed doublets have an anomaly.
Moreover, electrical charge neutrality has to be taken into account. A more careful
analysis shows \cite{Tu}, \cite{Ro} that in good approximation one ends up above $T_c$ with
\begin{equation}
B+L=-\frac{23}{79}(B-L)
\end{equation}
for the Standard Model with three generations and one Higgs doublet. (A mass correction
\cite{Ro} changes this slightly.) Without a primordial $B-L$ (e.g. from GUT physics) both
$B$ and $L$ would vanish. These considerations are relevant for scenarios of leptogenesis
(see \cite{DKu}).

\subsubsection*{Evolution below $T_c$}

The electroweak phase transition cannot generate the cosmological baryon asymmetry, because
the last of the following Sakharov conditions\footnote{ In the words of Sakharov:
``According to our hypothesis, the occurrence of $C$ asymmetry is the consequence of
violation of $CP$ invariance in the nonstationary expansion of the hot Universe during the
superdense stage, as manifest in the difference between the partial probabilities of the
charge conjugate reactions.''} is not fulfilled:

\vspace{0.5cm}

$\bullet$ Violation of baryon number

$\bullet$ Violation of $C$ and $CP$

$\bullet$ Deviations from thermodynamic equilibrium.

\vspace{0.5cm} The first two of these are obviously necessary. (The possibility that the
Universe is symmetric on scales much larger than the present horizon, but not in our
visible part, will be briefly discussed in the next section) The third condition requires
some explanation. A simple manner to understand its necessity is this: In thermodynamic
equilibrium one looses in a way the arrow of time (on cosmological short time scales),
hence the guaranteed $CTP$ symmetry implies $CP$ conservation. A more formal argument goes
as follows: In equilibrium the density matrix that maximizes the entropy for given average
values of the absolutely conserved `charges' $Q_i$ is the grand canonical ensemble
\[ \varrho = Z^{-1}\exp\Bigl[-\beta(H-\sum_i \mu_i Q_i)\Bigr] \]
($\mu_i= $chemical potentials for the conserved charges $Q_i$). Suppose that $B$ and $L$
are independently violated, and that beside  the electric charge $Q$ no other charges like
$B-L$ are conserved. Because of electrical charge neutrality $\varrho$ the has to be the
canonical ensemble: $\varrho=Z^{-1}e^{-\beta H}$. But $H$ is invariant under the $CPT$
operation $\theta$. Since $B$ is odd under $\theta$, we have
\[\langle B\rangle=Z^{-1}Tr(Be^{-\beta H})=Z^{-1}Tr\Bigl[\theta(Be^{-\beta H})\theta^{-1}\Bigr]
=-\langle B\rangle.\] So in thermodynamic equilibrium any previous baryonic and leptonic
asymmetries would be washed out. Note that such an argument could not be used above, since
not only $B-L$, but also all the anomaly free fermionic charges
$N_i-\frac{1}{3}B,~i=e,\mu,\tau$ (see (101) and (102)) are conserved. For this reason there
remain, after imposing charge neutrality, three\footnote{The differences $B_i-B_j$ for the
baryon numbers of the three generations are also anomaly free, however off-diagonal quark
interactions violate all $B_i$ (they preserve of course the total baryon number $B$).}
independent chemical potentials. That $B+L$ in (113) depends only on $B-L$ is the result of
the approximations used in \cite{Tu}, \cite{Ro}. Similarly, baryonic and leptonic
asymmetries may not be completely washed out if, beside the electric charge, there would
exist other absolutely conserved charges whose chemical potentials are non-vanishing. Below
$T_c$, as long as the `sphaleron processes' are sufficiently fast to establish thermal
equilibrium, the combination $B+L$ is no more proportional to $B-L$. As a result, even if
$B-L$ should vanish for all temperatures, a non-vanishing $B$ would result (mass effects
are important for this).

In a strongly first order transition there would be a departure from thermal equilibrium
and anomalous $B$ and $L$ violating processes could produce an asymmetry (with $B-L=0$).
Indeed, the transition would be very violent through nucleation of bubbles in the new phase
which would expand and collide. In order that the asymmetry is not erased after the phase
transition, the sphaleron energy at $T_c$ must be sufficiently high. This leads to the
following condition for the jump $v_T$ of the order parameter plotted in Fig. 7:
\begin{equation}
v_T/T_c>1.
\end{equation}
For a detailed discussion and references we refer to Sect. 7 of \cite{Sh}. The results
shown in Fig. 6 already imply that the electroweak transition in the Standard Model is too
weak for baryogenesis. As discussed earlier, the transition even stops to be of first order
for Higgs masses higher than $\simeq 72~GeV$, much below the LEP limit $m_H>110~GeV$.

In summary, the baryon asymmetry of the Universe must have been produced at energies much
higher than the electroweak scale. Part of the $B+L$ component will later be washed out by
anomalous weak processes, but the $B-L$ component would be preserved. Electroweak
baryosynthesis is only possible in extensions of the Standard Model. The matter-antimatter
asymmetry of the Universe is strong evidence for physics beyond the Standard Model.
Possible scenarios have been reviewed many times; see, e.g., \cite{Ri}, \cite{DKu}.

\subsection{QCD phase transition}

At temperatures of the order of the pion mass the pions are so abundant that they begin to
overlap. As a result, the interactions of their constituents (quarks and gluons) become
important.

For a rough estimate, consider the number density of a free pion gas in the early Universe
at $T\ll m_N$ (nucleon mass). We can take their chemical potential equal to zero, because
the pions are much more abundant than the nucleons. Their number density is for $T>m_\pi$:
\[ n_\pi(T)=3\int\frac{1}{e^{\varepsilon(p)/T} -1}\frac{d^3p}{(2\pi)^3}\simeq
\frac{3\zeta(3)}{\pi^2}T^3 .\] The volume $v_\pi$ of a pion is $(4\pi/3)r_\pi^3$, with the
pion radius $r_\pi\simeq 0.6~fm$. We find $n_\pi(T)v_\pi=1$ at a temperature
$T\simeq285~MeV$, with an energy density of $\simeq 850~MeV/fm^3$.

On the basis of the asymptotic freedom of QCD one naturally expects that at high
temperatures the hadronic matter behaves as an asymptotically free gas of quarks and gluons
(\textit{quark-gluon plasma}). Somewhere in the region $T\sim m_\pi$ there should be a
phase transition or crossover between the low energy hadronic matter -- with confinement
and broken chiral symmetry -- and the quark-gluon plasma. The study of the existence and
nature of this phase transition is a central theme of QCD. One is, of course, also
interested in the equation of state.

Because of serious infrared singularities, perturbation theory is of limited value, even
for the treatment of the asymptotic behavior of the quark-gluon plasma\footnote{For a brief
discussion of the severe infrared problems, see Chapter 20 of Ref. \cite{Rot}. Even
resummed perturbation theory does not cure the infrared problems in QCD.}. This is even
more so when the phase transition is addressed. Therefore, the non-perturbative lattice
approach is the only way to investigate the thermodynamics of hadrons from first
principles. A lot of work has been invested in this domain over many years. Very useful
general references are \cite{MM}, \cite{Ka} and \cite{Go}. Since the critical temperature
is close to the strange quark mass, it is important to simulate dynamical light quark
flavors realistically. This has almost been achieved by now. A recent study of the MILC
collaboration \cite{He} comes to the conclusion that in the real world there is presumably
\textit{no bona fide phase transition at the physical quark masses}. In Fig. 8 we reproduce
the result for the order parameter $\langle\bar{\psi}\psi\rangle$ for two degenerate light
quarks (up and down) plus the strange quark with its physical mass.

\begin{figure}
\begin{center}
\includegraphics[height=0.6\textheight]{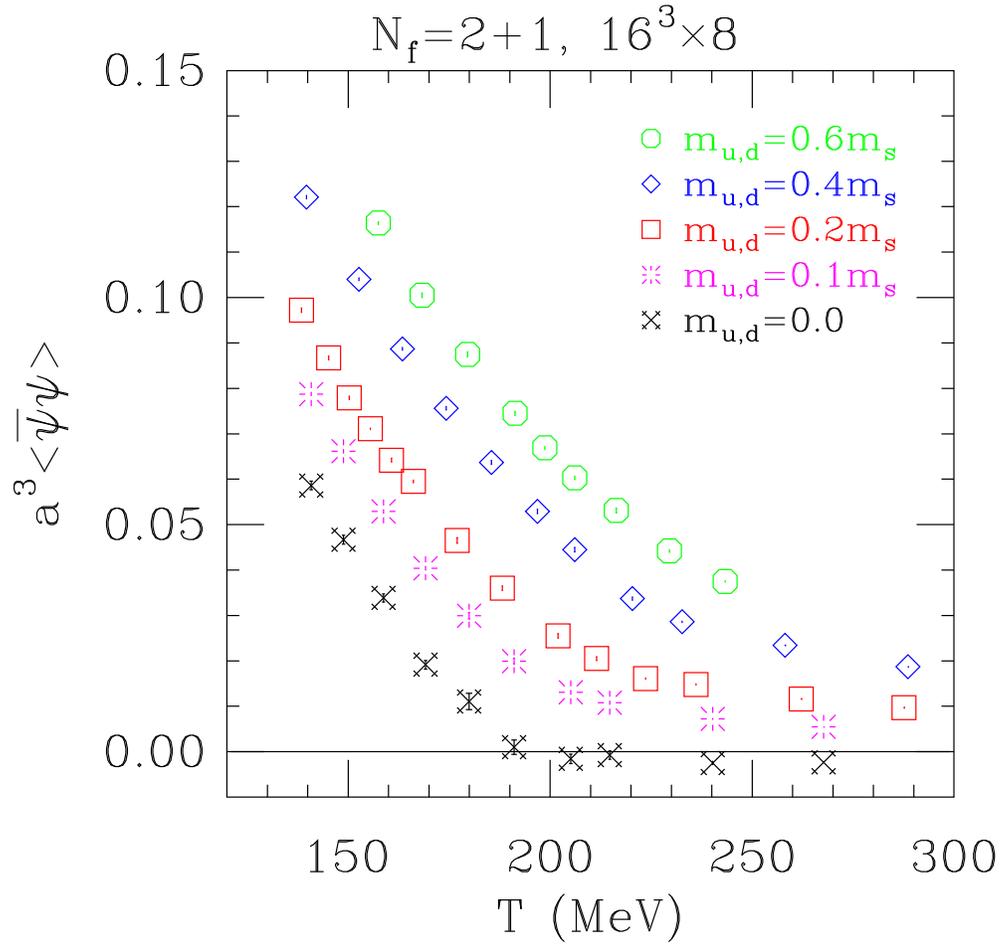}
\caption{The chiral order parameter, $\bar{\psi}\psi$, on $16^3\times 8$ lattices (from
Ref. \cite{He}).} \label{Fig-8}
\end{center}
\end{figure}

Fig. 9 shows the real part of the \textit{Wilson line} or \textit{Polyakov loop}, defined
as (using the notation $U_{\mu}(x)$ for the link variables)
\begin{equation}
L(\mathbf{x})=Tr\left\{\prod_{x_4=1}^{N_t}U_4(\mathbf{x},x_4)\right\},
\end{equation}
with $N_t=\beta /a$. This quantity is gauge invariant. The logarithm of its expectation
value is proportional to the free energy of the system with a single heavy quark, measured
relative to that in the absence of such a quark. $\langle L\rangle =0$ can be interpreted
as signalizing confinement.

\begin{figure}
\begin{center}
\includegraphics[height=0.6\textheight]{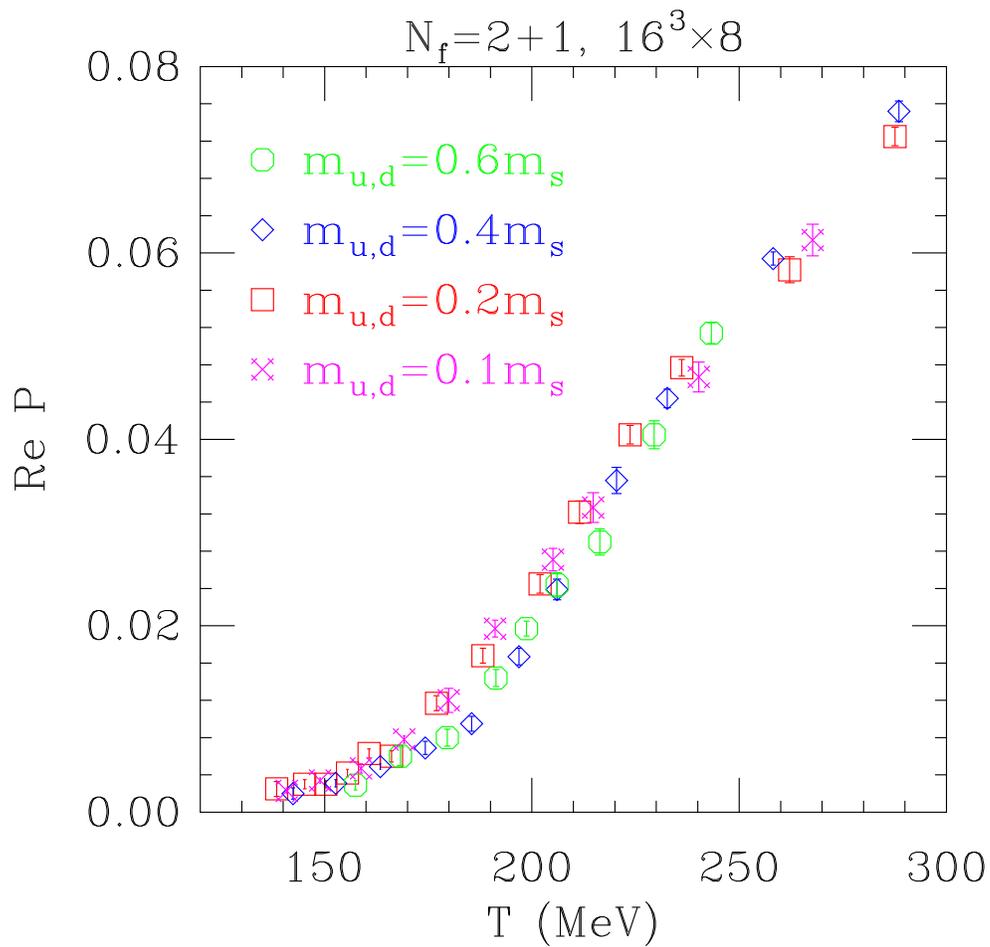}
\caption{The real part of the Polyakov loop on $16^3\times 8$ lattices (from Ref.
\cite{He}).} \label{Fig-9}
\end{center}
\end{figure}

These and other results suggest that there is a crossover, rather than a sharp phase
transition from the confined to the unconfined behavior. Similar results were obtained in
\cite{K2}.

Many cosmological speculations connected with the QCD transition (possibility of
inhomogeneous nucleosynthesis, strange quark nuggets, etc) have been discussed. Since it is
not clear whether there are any observable footprints of this transition the Universe
passed through (and for limitations of time), I leave this interesting subject.

In the central regions of neutron stars matter is squeezed to densities several times
higher than the nuclear density, $\rho_{nuc}=2.8\times 10^{14}~g~cm^{-3}$ (see, e.g.,
Chap.6 of \cite{NS}). One expects that these regions are, at least for the most massive
neutron stars, in the deconfined chirally symmetric phase. It is also expected that at
relatively low temperatures the transition to this state of matter as a function of baryon
chemical potential $\mu_B~(=3\mu_{quark})$ is first order for realistic quark masses. It is
difficult to perform lattice simulations for $\mu_B\neq0$, because the quark determinant
(arising from the functional integral over the quark fields) becomes \textit{complex}, and
thus also the corresponding term in the effective action. For a clear discussion of these
difficulties we refer to Sect.5.4.3 of \cite{MM}. Recently, significant progress has been
made in algorithms, and realistic calculations may become feasible.

\subsection{Cosmic topological defects}

The formation of topological defects is well-known to condensed matter physicists. A
classic example is the formation of vortices in type II superconductors. As you all know
these are described by the Ginzburg-Landau theory, which is a `spontaneously broken' gauge
theory (the Abelian Higgs model). The complex order parameter $\phi(x)$, which later turned
out to be proportional to the gap function in BCS theory, plays in particle physics
language the role of a Higgs field. I recall that the integer number in the flux
quantization is also the degree (winding number) of the map which associates to each
direction in the two-dimensional transverse space (topologically a circle $S^1$) the
asymptotic value of $\phi$ in the vacuum manifold. The latter is the circle on which
$|\phi|$ assumes the minimum of the Ginzburg-Landau (Higgs) potential.

The analogous 't Hooft-Polyakov monopoles exist, for example as solutions of the
$SU(2)$-Higgs model. The pole strength is then again topologically quantized, and the
corresponding integer number is this time the winding number of the map which associates to
each direction of 3-space (topologically $S^2$) the asymptotic value of $\phi$ in the
vacuum manifold, that is this time an $S^2$. Note that the gauge group $SU(2)$acts on this
vacuum manifold transitively and is thus topologically isomorphic to the homogeneous space
$SU(2)/H$, where $H=U(1)$ is the stabilizer of any point of the vacuum manifold. For a
general gauge theory with gauge group $G$, non-Abelian (regular) monopoles exist if the
second homotopy group of $G/H~ (H$ is again the stabilizer of the vacuum manifold) is
non-trivial. Homotopy theory allows to compute these homotopy groups quite easily.

Textures are also known in condensed matter physics. These are classified by a third
homotopy group.

The theory of topological defects in cosmology is extensively described in the monograph
\cite{Vi}. On the basis of what we now know about the cosmic microwave background, the
conclusion is inevitable that topological defects played no important role in large scale
structure formation. This is the main reason why interest in the subject has declined.

\section{Vacuum energy problem and Dark Energy}

In the course of the various `phase' transitions the Universe passed through, the free
energy density changed by a huge amount in comparison to the present day cosmological
energy density (which is close to the critical value). This gives me the opportunity to
discuss the \textit{cosmological constant problem}, a profound mystery indeed of present
day physics.

\subsection{ Vacuum energy and gravity}

When we consider the coupling to gravity, the vacuum energy density acts like a
cosmological constant. In order to see this, first consider the vacuum expectation value of
the energy-momentum tensor in Minkowski spacetime. Since the vacuum state is Lorentz
invariant, this expectation value is an invariant symmetric tensor, hence proportional to
the metric tensor. For a curved metric this is still the case, up to higher curvature
terms:
\begin{equation}
<T_{\mu\nu}>_{vac} = g_{\mu\nu}\rho_{vac} + \textit{higher curvature terms}.
\end{equation}
The {\it effective} cosmological constant, which controls the large scale behavior of the
Universe, is given by
\begin{equation}
\Lambda = 8\pi G\rho_{vac}+\Lambda_{0},
\end{equation}
where $\Lambda_0$ is a bare cosmological constant in Einstein's field equations.

We know from astronomical observations since a long time that
$\rho_\Lambda\equiv\Lambda/8\pi G$ can not be larger than about the critical density:
\begin{eqnarray}
\rho_{crit} &=& \frac{3H_0^2}{8\pi G} \nonumber \\
&=& 1.88\times 10^{-29} h_0^2 ~\textit {g}~\textit {cm}^{-3} \\
&\simeq & (3\times10^{-3}eV)^4, \nonumber
\end{eqnarray}
where $h_0$ is the {\it reduced Hubble parameter}
\begin{equation}
h_0 = H_0/(100~\textit{km}~\textit{s}^{-1}\textit{Mpc}^{-1})
\end{equation}
and is close to 0.7.

It is a complete mystery as to why the two terms in (117) should almost exactly cancel.
This is -- more precisely stated -- the famous $\Lambda$-problem.

As far as I know, apart some unpublished remarks of Pauli in the early 1920's \cite{NS2},
the first who wondered about possible contributions of the vacuum energy density to the
cosmological constant was Zel'dovich . He discussed this issue in two papers \cite{Ze}
during the third renaissance period of the $\Lambda$-term, but before the advent of
spontaneously broken gauge theories. The following remark by him is particularly
interesting. Even if one assumes completely ad hoc that the zero-point contributions to the
vacuum energy density are exactly cancelled by a bare term, there still remain higher-order
effects. In particular, {\it gravitational} interactions between the particles in the
vacuum fluctuations are expected on dimensional grounds to lead to a gravitational
self-energy density of order $G\mu^6$, where $\mu$ is some cut-off scale. Even for $\mu$ as
low as 1 GeV (for no good reason) this is about 9 orders of magnitude larger than the
observational bound.

This illustrates that there is something profound that we do not understand at all,
certainly not in quantum field theory ( so far also not in string theory).  We are unable
to calculate the vacuum energy density in quantum field theories, like the Standard Model
of particle physics. But we can attempt to make what appear to be reasonable
order-of-magnitude estimates for the various contributions. \textbf{All expectations} (some
of which are discussed below) are \textbf{in gigantic conflict with the facts.} Trying to
arrange the cosmological constant to be zero is unnatural in a technical sense. It is like
enforcing a particle to be massless, by fine-tuning the parameters of the theory when there
is no symmetry principle which implies a vanishing mass. The vacuum energy density is
unprotected from large quantum corrections. This problem is particularly severe in field
theories with spontaneous symmetry breaking. In such models there are usually several
possible vacuum states with different energy densities. Furthermore, the energy density is
determined by what is called the effective potential, and this is {\it dynamically}
determined. Nobody can see any reason why the vacuum of the Standard Model we ended up as
the Universe cooled, has -- for particle physics standards --  an almost vanishing energy
density. Most probably, we will only have a satisfactory answer once we shall have a theory
which successfully combines the concepts and laws of general relativity about gravity and
spacetime structure with those of quantum theory.

\subsection{Simple estimates of vacuum energy contributions}

If we take into account the contributions to the vacuum energy from vacuum fluctuations in
the fields of the Standard Model up to the currently explored energy, i.e., about the
electroweak scale $M_F = G_F^{-1/2}\approx 300~ GeV ~ (G_F:$ Fermi coupling constant), we
cannot expect an almost complete cancellation, because there is {\it no symmetry principle}
in this energy range that could require this. The only symmetry principle which would imply
this is {\it supersymmetry}, but supersymmetry is broken (if it is realized in nature).
Hence we can at best expect a very imperfect cancellation below the electroweak scale,
leaving a contribution of the order of $M_F^4$ . (The contributions at higher energies may
largely cancel if supersymmetry holds in the real world.)

We would reasonably expect that the vacuum energy density is at least as large as the
condensation energy density of the QCD  transition to the confined  phase of broken chiral
symmetry. Already this is far too large, namely of the order $ \sim \Lambda_{QCD}^4/
16\pi^2 \sim 10^{-4} \textit{GeV}^4$, i.e., {\it more than 40 orders of magnitude larger}
than $\rho_{crit}$. Beside the formation of quark condensates $<\bar{q}q>$ in the QCD
vacuum which break chirality, one also expects a gluon condensate $<G^{\mu\nu}_{a}
G_{a\mu\nu}> \sim \Lambda_{QCD}^4.$ This produces a significant vacuum energy density as a
result of a dilatation anomaly: If $\Theta_\mu^\mu$ denotes the ``classical'' trace of the
energy-momentum tensor, we have \cite{Ca}
\begin{equation}
T^\mu_\mu = \Theta^\mu_\mu + \frac{\beta(g_3)}{2g_3} G^{\mu\nu}_{a}G_{a\mu\nu},
\end{equation}
where the second term is the QCD piece of the trace anomaly ($\beta(g_3)$ is the
$\beta$-function of QCD that determines the running of the strong coupling constant). I
recall that this arises because a scale transformation is no more a symmetry if quantum
corrections are included. Taking the vacuum expectation value of (120), we would again
naively expect that $<\Theta_\mu^ \mu>$ is of the order $M_F^4$. Even if this should vanish
for some unknown reason, the anomalous piece is cosmologically gigantic. The expectation
value $<G^{\mu\nu}_a G_{a\mu\nu}>$ can be estimated with QCD sum rules \cite{SS}, and gives

\begin{equation}
<T^{\mu}_\mu>^{anom}\sim (350 MeV)^4,
\end{equation}
about 45 orders of magnitude larger than $\rho_{crit}$. The expected energy density is
larger than the energy-mass density in a neutron star. This reasoning should show
convincingly that the cosmological constant problem is indeed a profound one. (Note that
there is some analogy with the (much milder) strong CP problem of QCD. However, in contrast
to the $\Lambda$-problem, Peccei and Quinn \cite{PQ} have shown that in this case there is
a way to resolve the conundrum.)

Let us also have a look at the Higgs condensate of the electroweak theory. Recall that in
the Standard Model we have for the Higgs doublet $\Phi$ in the broken phase for
$<\Phi^{*}\Phi>\equiv\frac{1}{2}\phi^2$ the potential
\begin{equation}
V(\phi) = -\frac{1}{2}m^2\phi^2 + \frac{\lambda}{8}\phi^4.
\end{equation}
Setting as usual $\phi=v+H$, where $v$ is the value of $\phi$ where $V$ has its minimum,
\begin{equation}
v= \sqrt{\frac{2m^2}{\lambda}} = 2^{-1/4}G_F^{-1/2}  \sim 246 GeV,
\end{equation} we find that the Higgs mass is related to $\lambda$ by
$\lambda= M_{H}^2/v^2$. For  $\phi=v$ we obtain the energy density of the Higgs condensate
\begin{equation}
V(\phi=v)=-\frac{m^4}{2\lambda}= -\frac{1}{8\sqrt{2}}M_F^2M_H^2= \mathcal{O}(M_F^4).
\end{equation}
We can, of course, add a constant $V_0$ to the potential (122) such that it cancels the
Higgs vacuum energy in the broken phase -- including higher order corrections. This again
requires an extreme fine tuning. A remainder of only $\mathcal{O}(m_e^4)$, say, would be
catastrophic. This remark is also highly relevant for models of inflation and quintessence.

In attempts beyond the Standard Model the vacuum energy problem so far remains, and often
becomes even worse. For instance, in supergravity theories with spontaneously broken
supersymmetry there is the following simple relation between the gravitino mass $m_g$ and
the vacuum energy density
\[ \rho_{vac} = \frac{3}{8\pi G}m_g^2.\]
Comparing this with eq.(30) we find
\[ \frac{\rho_{vac}}{\rho_{crit}} \simeq 10^{122}\Bigl (\frac{m_g}{m_{Pl}}\Bigr )^2.\]
Even for $m_g \sim 1\; eV$ this ratio becomes $10^{66}$. ($m_g$ is related to the parameter
$F$ characterizing the strength of the supersymmetry breaking by $m_g = (4\pi G/3)^{1/2}
F$, so $m_g\sim 1\; eV$ corresponds to $F^{1/2} \sim 100\; TeV$.)

Also string theory has not yet offered convincing clues why the cosmological constant is so
extremely small. The main reason is that a {\it low energy mechanism} is required, and
since supersymmetry is broken, one again expects a magnitude of order $M_F^4$, which is
{\it at least 50 orders of magnitude too large} (see also \cite{Wi}). However,
non-supersymmetric physics in string theory is at the very beginning and workers in the
field hope that further progress might eventually lead to an understanding of the
cosmological constant problem.

I hope I have convinced you, that there is something profound that we do not understand at
all, certainly not in quantum field theory, but so far also not in string theory.

For more on this, see e.g. \cite{NS2} (and references therein).

\subsection{Astronomical evidence for Dark Energy}

A wide range of astronomical data support the following `concordance' $\Lambda$- Cold-
Dark- Matter ($\Lambda$CDM) model: The Universe is spatially flat and dominated by vacuum
energy density or an effective equivalent, and weakly interacting cold dark matter.
Furthermore, the primordial fluctuations are adiabatic and nearly scale invariant, as
predicted in simple inflationary models. It is very likely that the present concordance
model will survive. The evidence for a dark energy component of about 0.7 times the
critical energy density is steadily increasing, and by the time the proceedings of this
school will be available one can presumably make stronger statements. Readable reviews (for
people outside the field) of the current evidence for a nearly homogeneous energy density
with negative pressure that dominates the energy content of the recent and future Universe
are \cite{Hu}, \cite{Dod}. Beside, there is a rapidly growing semi-popular literature on
this exciting subject, confronting us with really profound fundamental physics questions.

\section*{Acknowledgements}

I would like to thank the Organizers of the Zuoz Summer School for the invitation to this
beautiful place. Discussions with participating experts on phase transitions have been very
useful. Special thanks go to Walter Fischer, the initiator of this sequence of summer
schools. I am grateful to Urs Heller, who sent me the updated figures 8 and 9 in these
lecture notes. Several remarks by Andreas Wipf on an early version of the manuscript have
been helpful. Finally, I would like to thank Mikko Laine for comments on the first version
on astro-ph, concerning the endpoint of the $1^{st}$ order transition in the electroweak
theory.

\end{document}